\documentclass[manuscript]{aastex}
\usepackage{fancyhdr}
\usepackage{longtable}
\usepackage{latexsym}
\usepackage{graphicx}
\usepackage{amsmath}
\usepackage{natbib}
\usepackage{upgreek}
\usepackage{multirow}
\usepackage{array}
\usepackage{pdflscape}
\usepackage{graphicx,subfigure}
\usepackage{tabularx}
\usepackage{epsfig}
\usepackage{color}
\usepackage{lscape}
\usepackage{ulem}

\newcommand{\PreserveBackslash}[1]{\let\temp=\\#1\let\\=\temp}
\newcolumntype{C}[1]{>{\PreserveBackslash\centering}p{#1}}
\newcolumntype{R}[1]{>{\PreserveBackslash\raggedleft}p{#1}}
\newcolumntype{L}[1]{>{\PreserveBackslash\raggedright}p{#1}}

\shorttitle{Applicaition of SSRC to LAMOST Calibration}
\shortauthors{B. Du et al.}

\begin{document}
%\begin{CJK}{UTF8}{gbsn}

\title{LAMOST Spectrograph Response Curves: Stability and  Application to flux calibration}

\author{Bing Du\altaffilmark{1,2} ,
       A-Li Luo \altaffilmark{1,2},
       Zhong-Rui Bai \altaffilmark{1,2},
       Xiao Kong \altaffilmark{1},
       Jian-Nan Zhang \altaffilmark{1},
       Yan-Xin Guo \altaffilmark{1},
       Neil James Cook  \altaffilmark{3},
       Wen Hou  \altaffilmark{1},
       Hai-Feng Yang \altaffilmark{1},
       Yin-Bi Li \altaffilmark{1},
       Yi-Han Song  \altaffilmark{1},
       Jian-Jun Chen \altaffilmark{1},
       Fang Zuo \altaffilmark{1},
       Ke-Fei Wu \altaffilmark{1},
       Meng-Xin Wang \altaffilmark{1},
      You-Fen Wang \altaffilmark{1}, and
      Yong-Heng Zhao \altaffilmark{1}
}

\altaffiltext{1}{Key Laboratory of Optical Astronomy, National Astronomical Observatories, Chinese Academy of Sciences, Beijing 100012, China. \sf lal@nao.cas.cn}
\altaffiltext{2}{University of Chinese Academy of Sciences, Beijing 100049, China.}
\altaffiltext{3}{Centre for Astrophysics Research, School of Physcis, Astronomy and Mathematics, University of Hertfordshire, College Lane, Hartfield AL10 9AB, UK.}

\begin{abstract}
The task of flux calibration for LAMOST (Large sky Area Multi-Object Spectroscopic Telescope) spectra is difficult due to many factors. For example, the lack of  standard stars, flat fielding for large field of view, and variation of reddening between different stars especially at low galactic latitudes etc. Poor selection, bad spectral quality, or extinction uncertainty of standared  stars not only might induce errors to the calculated spectral response curve (SRC), but also might lead to failures in producing final 1D spectra. In this paper,  we inspected spectra with  Galactic latitude $|b|\geq 60^{\circ}$  and reliable stellar parameters,  determined through the LAMOST Stellar Parameter Pipeline (LASP), to study the stability of the spectrograph. To guarantee the selected stars had been observed by each fiber, we selected 37,931 high quality exposures of 29,000 stars from LAMOST DR2, and more than 7 exposures for each fiber. We calculated the SRCs for each fiber for each exposure, and calculated the statistics of SRCs for spectrographs with both the fiber variations and time variations. The result shows that the average response curve of each spectrograph (henceforth ASPSRC) is relatively stable with statistical errors $\leq$ 10\%. From the comparison between each ASPSRC and the SRCs for the same spectrograph obtained by 2D pipeline, we find that the ASPSRCs are good enough to use for the calibration. The ASPSRCs have been applied to spectra which were abandoned by LAMOST 2D pipeline due to the lack of standard stars, increasing the number of LAMOST spectra by 52,181 in DR2.  Comparing those same targets with SDSS,  the relative flux differences between SDSS spectra and that of LAMOST spectra with the ASPSRC method are less than 10\%, which underlines that the ASPSRC method is feasible for LAMOST flux calibration.

\end{abstract}


\section{Introduction}

The LAMOST is a quasi--meridian reflecting Schmidt telescope with an effective aperture of $\sim$4m and field of view (FoV) of 5 degree in diameter.  At the focal plane,  4,000 robotic optical fibers of aperture size 3.3 arcsec projected on the sky relay the target light to 16 spectrographs,  250 fibers each \citep{cui,deng}.  Proceeded by one-year Pilot Survey,  the LAMOST Regular Surveys started in September 2012.

The wavelength range of LAMOST covers  3,700 to 9,000 \AA \ and is recorded in two arms, a blue arm (3,700-5,900 \AA) and a red arm (5,700-9,000 \AA), with the resolving power of R$\sim$1,800.  A final spectrum is obtained by merging several exposures and connecting wavelength bands.

Raw data from the LAMOST surveys are reduced with the LAMOST 2D pipeline \citep{lal}. The procedures used by the 2D pipeline, similar to those of SDSS \citep{Stoughton}, aim to extract spectra from the CCD images and then calibrate them. The main tasks of the 2D pipeline include the steps of fiber tracing, flux extraction, wavelength calibration, flat fielding, sky subtraction, flux calibration, multi-exposure co--addition and the connection of the two wavelength bands. Since the data reduction steps are the reverse process of the data acqusition process, we should understand the  data acquisition process of LAMOST, which can be simplified as follows.
\begin{equation}
  F_{o}(j,\lambda)=[F_i(j,\lambda)\times d_s(\lambda)+sky_r(\lambda)]\times d_f(\lambda) \times d_p(\lambda)+scatter(j,\lambda)+C_k(j,\lambda)+B
\label{eq:eq1}
\end{equation}
In this equation, $F_{o}(j,\lambda)$ is the observed signal, where $\it{j}$ denotes the $\it{j}$--th fiber and $\lambda$ denotes the wavelength;

$F_i(j,\lambda)$ is the target signal before pass through the atmosphere;

$d_s(\lambda)$ is the extinction function including atmospheric and interstellar reddening;

$sky_r(\lambda)$ is sky background;

$d_f(\lambda)$ is the fiber transmission function, a random number selected from a Gaussian distribution, with a mean of 0.9 and variance of 1.0;

$d_p(\lambda)$ is the spectral response function due to the dispersion of the spectrograph;

$scatter(j,\lambda)$ is the scattering light including symmetrical scattering and the cross-contamination of fibers;

$C_k(j,\lambda)$ is the parameter to compensate with cosmic rays;

$B$ is the CCD background.

The purpose of the LAMOST flux calibration is to remove the spectral response curve (SRC) from observations. Considering that $d_f(\lambda)$ is divided during the flat field, the SRCs of spectrographs could be simplified as shown in equation \ref{eq:eq2}, which only includes $d_s(\lambda)$ and $d_p(\lambda)$.
\begin{equation}
  SRC(j,\lambda)=d_s(\lambda) \times d_p(\lambda)
\label{eq:eq2}
\end{equation}

In the real flux calibration process, $d_s(\lambda)$ and $d_p(\lambda)$  are considered as a whole SRC, by which the single exposure is divided. For the LAMOST 2D pipeline, selection of standard stars is the first step of flux calibration \citep{Song}. The pipeline selects standard stars automatically by comparing all the observed spectra with the KURUCZ library produced based on the ATLAS9 stellar atmosphere model data \citep{castelli}.  For each of the 16 spectrographs, several high quality spectra with SNR$\geq$10,  5,750 K$\leq T_{\rm eff} \leq$7,250 K, \ log g$\geq$3.5 dex and -1.0 dex$\leq$[Fe/H]$\leq$ 0 dex are selected as standard stars. Actually, the LAMOST 2D pipeline picks out standards with the temperature in the range of 6,000--7,000 K at first step, if there is not enough stars in this range, the 2D pipeline will extend the range to 5,750--7,250 K.  If more than 3 standard stars are found for a spectrograph, the SRCs of the spectrograph can be derived by comparing the observed spectra with synthetic spectra (using the corresponding parameters from the KURUCZ spectral library). Because the 2D pipeline estimates the parameters by simple fitting  with KURUCZ model, the parameters have great uncertainties for the stars with [Fe/H]$<$-1.0 dex, meanwhile considering that the number of metal poor stars is small in each spectrograph, the 2D pipeline uses the metallicity cut of -1.0 dex$\leq$[Fe/H]$\leq$ 0 dex for the selection of the standards.  Unfortunately, for the current LAMOST 2D pipeline, when there are not enough suitable standard stars, data reduced of the spectrograph  has to be suspended.

In this paper, to rescue the unsuccessful spectra, we propose a novel flux calibration method based on the stability analysis of the SRCs. Thanks to more than 2 million spectra, with reliable stellar parameters in DR2,  we are able to statistically measure the instrument stability. Through stellar parameters, the SRC of each fiber could be obtained. By averaging SRCs in each spectrograph, we can get an average spectrograph SRC (ASPSRC), and use it to calibrate spectra in each spectrograph without pre-selecting the flux standard stars assuming the ASPSRC is sufficiently stable. This flux calibration method can rescue more spectra from LAMOST which were abandoned by 2D pipeline.

The paper is organized as follows. Section 2 details of the procedures  used to create the ASPSRC  for each spectrograph. The accuracy analysis of  the ASPSRC and its application to flux calibration are presented in Section 3.  We conclude with Section 4  which summarizes and discusses the results.

\section{Statistical Spectrograph Response Curves}

\subsection{Selection of the Sample}

Work by Xiang et al.(2015) show that variations of the SRCs exist, this is done by using stars in high dense fields, however these suffer from high interstellar extinction. However,  to study the variations of the SRCs one should use stars with less extinction.  Therefore, we selected stars at high Galactic latitude to analyze instrument response \citep{Fitz1,Fitz2}.

To obtain a good approximation of the ASPSRCs, we require as many flux standard stars possible. To ensure the quality of the sample, the stellar parameters of LASP \citep{wu,wu01} were used to select the F--stars with the highest signal to noise ratios (SNRs). We selected stars with 6,000 K$\leq T_{\rm eff} \leq$7,000 K, log g$\geq$3.5 dex and Galactic latitude $| b | \geq 60^{\circ}$.  This resulted in 29,526 targets with 37,931 pairs (blue and red arm ) of single exposures. For the ASPSRC, we have enough stars in the temperature range 6,000 K$\leq T_{\rm eff} \leq$7,000 K, so we don't need to extend to 5,750 K$\leq T_{\rm eff} \leq$7,250 K as the 2D pipeline does.  Metallicity mainly affect the blue--arm spectra at wavelengths less than 4,500 \AA.  An error of 0.2 dex in [Fe/H] can change the SED shape between 3, 800 \AA \ and 4,500 \AA \ by approximately 3\%, while the effects at wavelengths greater than 4,500 \AA  \ are only marginally \citep{xiang}. As the ASPSRCs are derived from a great number of standard stars instead of a group of several standards in the 2D pipeline, and 90\% of the metallicities are in the range of  [Fe/H]$\geq$ -1.0 dex, which the averaged SRCs are generated from. The accuracy of the parameters measured by LASP are good enough even for metal poor stars, and will not affect the averaged result. Thus we did not use a metallicity cut in this sample selection.

\begin{figure}
\centering
\includegraphics[width=12cm]{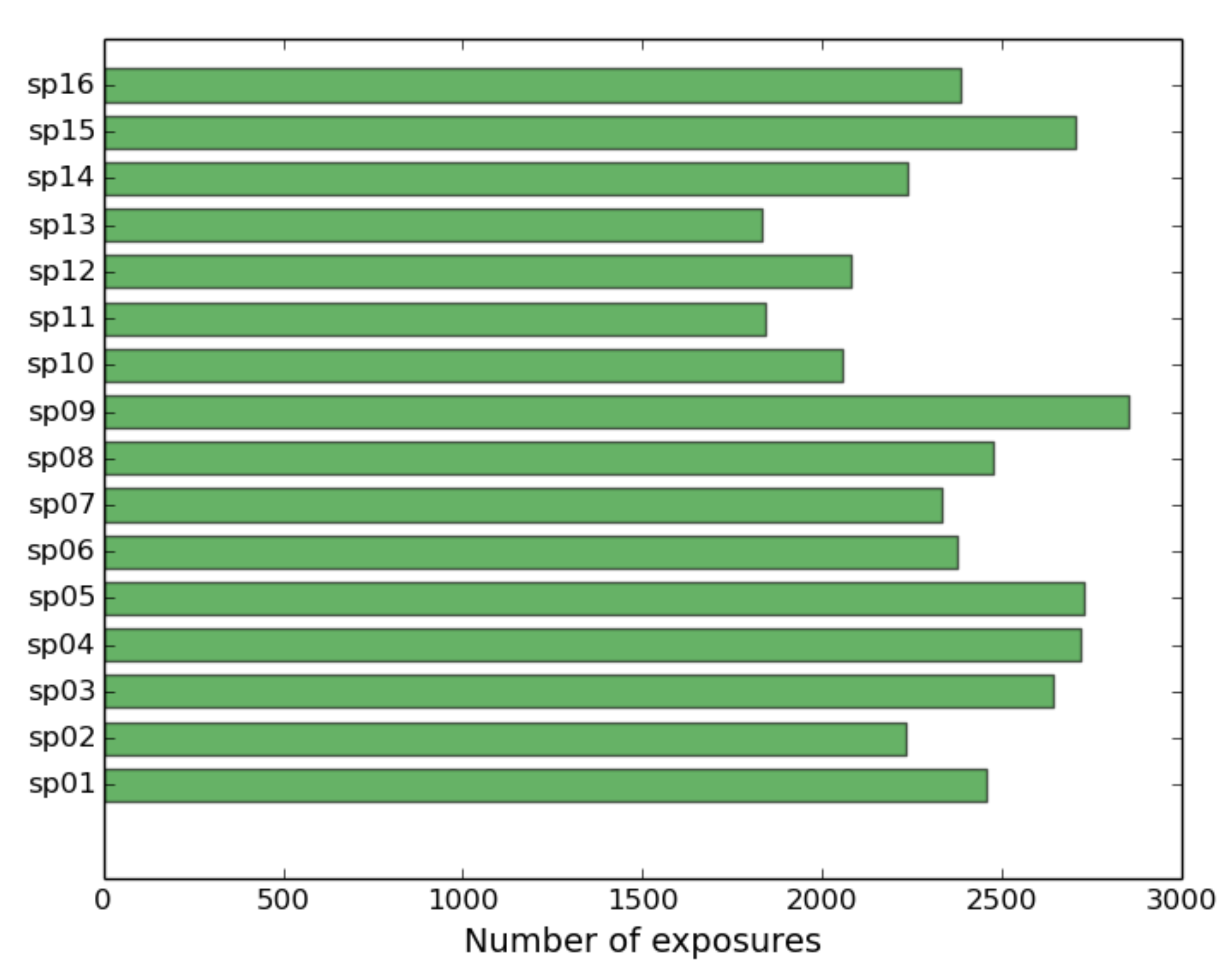}
\caption{Histogram of the numbers of exposures of standard stars selected per spectrograph.}
\label{fig1}
\end{figure}

\begin{figure}
\centering
\includegraphics[width=12cm]{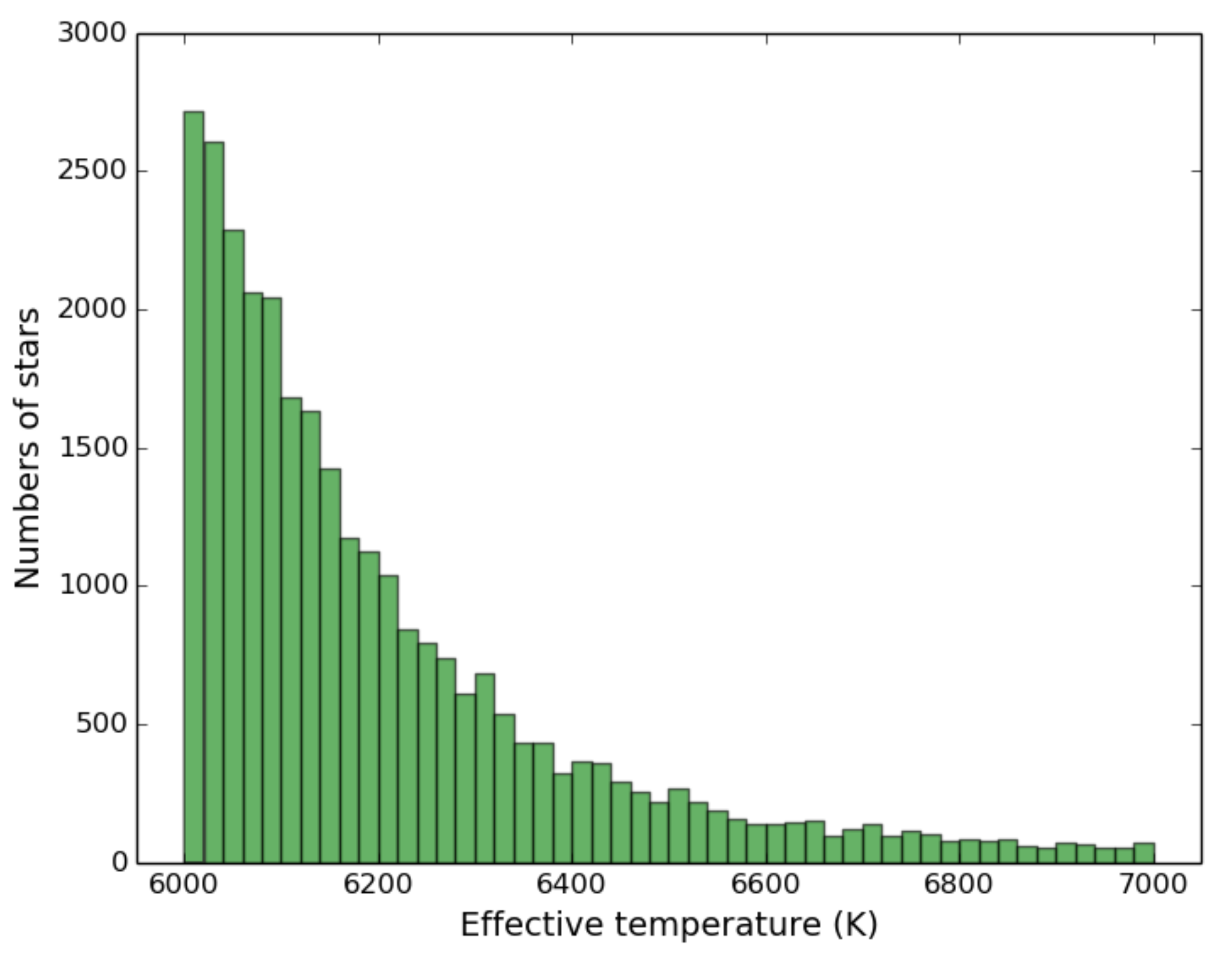}
\caption{\textbf{Histogram} of the effective temperatures of the selected standard stars.}
\label{fig2}
\end{figure}

With the benefit of the large sample of targets that satisfied the above parameter space, we find that there are sufficient and appropriate exposures across all fibers and spectrographs to allow us to use them as standards. Fig \ref{fig1} shows the histogram of the number of standards per spectrograph from DR2, which indicates that  at least 7 standards per fiber (with 250 fibers in each spectrograph this is equivalent to at least 1,750 individual exposures, shown in Fig \ref{fig1}). Fig \ref{fig2} shows the histogram of their effective temperatures, mostly located in the vicinity of 6,100 K (i.e. F8 type stars).

\subsection{Spectral Response Curves}

Let $F_o(\lambda)$ and $F_i(\lambda)$ denote the measured and intrinsic spectral flux density \textbf{thus},
\begin{equation}
  F_o(\lambda)=d_s(\lambda)d_p(\lambda)F_i(\lambda)
\label{eq:eq3}
\end{equation}

where $d_s(\lambda)$ is the combined atmospheric and interstellar extinction, and $d_p(\lambda)$ the telescope and instrumental response.  In this work, we adopted synthetic flux as $F_i(\lambda)$, which is calculated using SPECTRUM based on the ATLAS9 stellar atmosphere model data released by Fiorella Castelli.  The synthetic spectra of  1\AA \ dispersion from the library of KURUCZ were used, then the spectra were degraded to the LAMOST spectral resolution by convolving with a Gaussian function. Only those  with  a constant micro--turbulent velocity of 2.0 km/s and a zero rotation velocity were adopted, since these two parameters have little effect on the spectral energy distribution (SED) at a given temperature \citep{grupp}.

The interstellar extinction can be neglected owing to our selection of high latitude standards, however the atmospheric extinction can not be separated from instrumental response. The SRCs in this paper include atmospheric extinction, and their variations are included in the uncertainty of the SRCs.

\begin{figure}
\includegraphics[width=8.0cm]{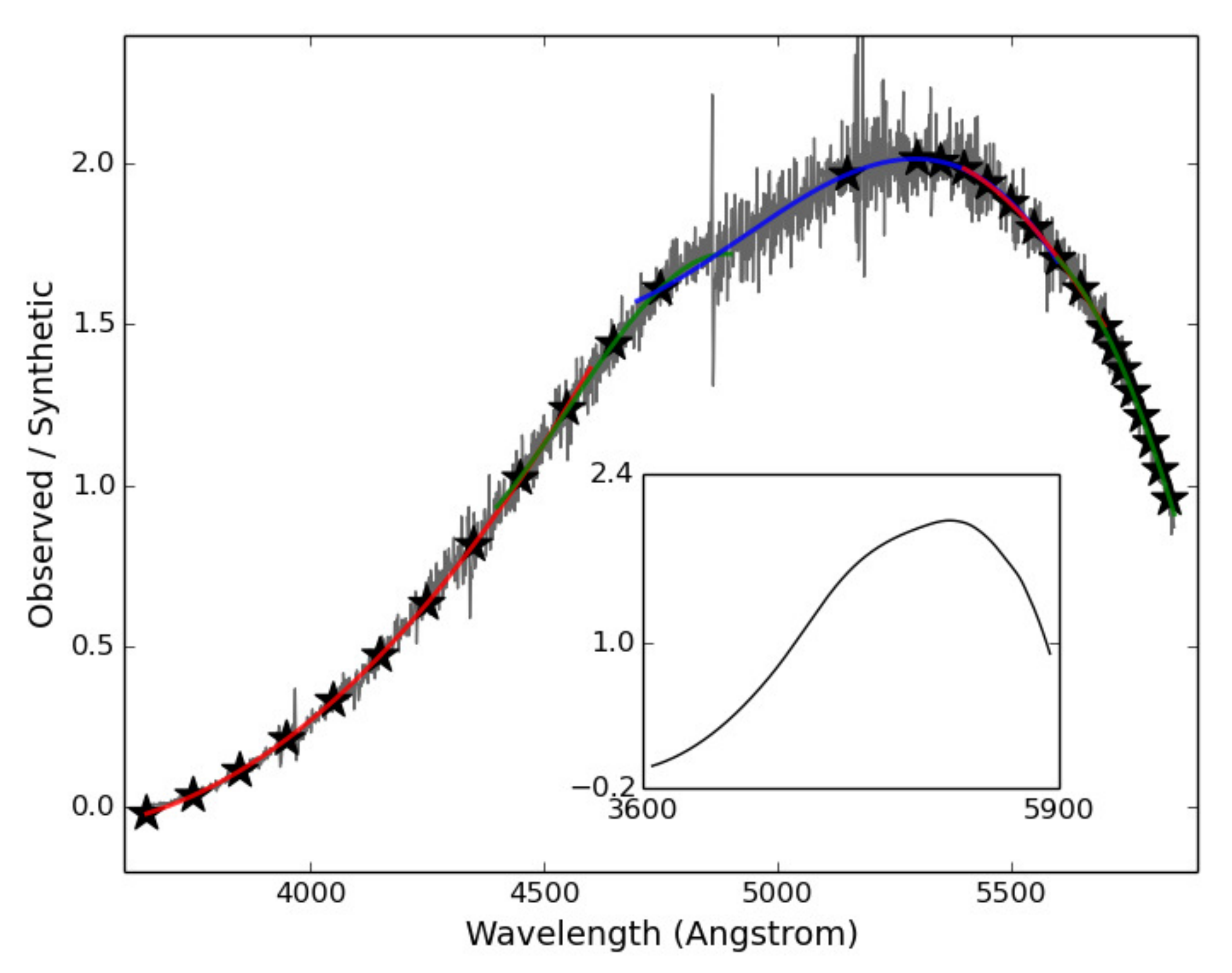}
\includegraphics[width=8.0cm]{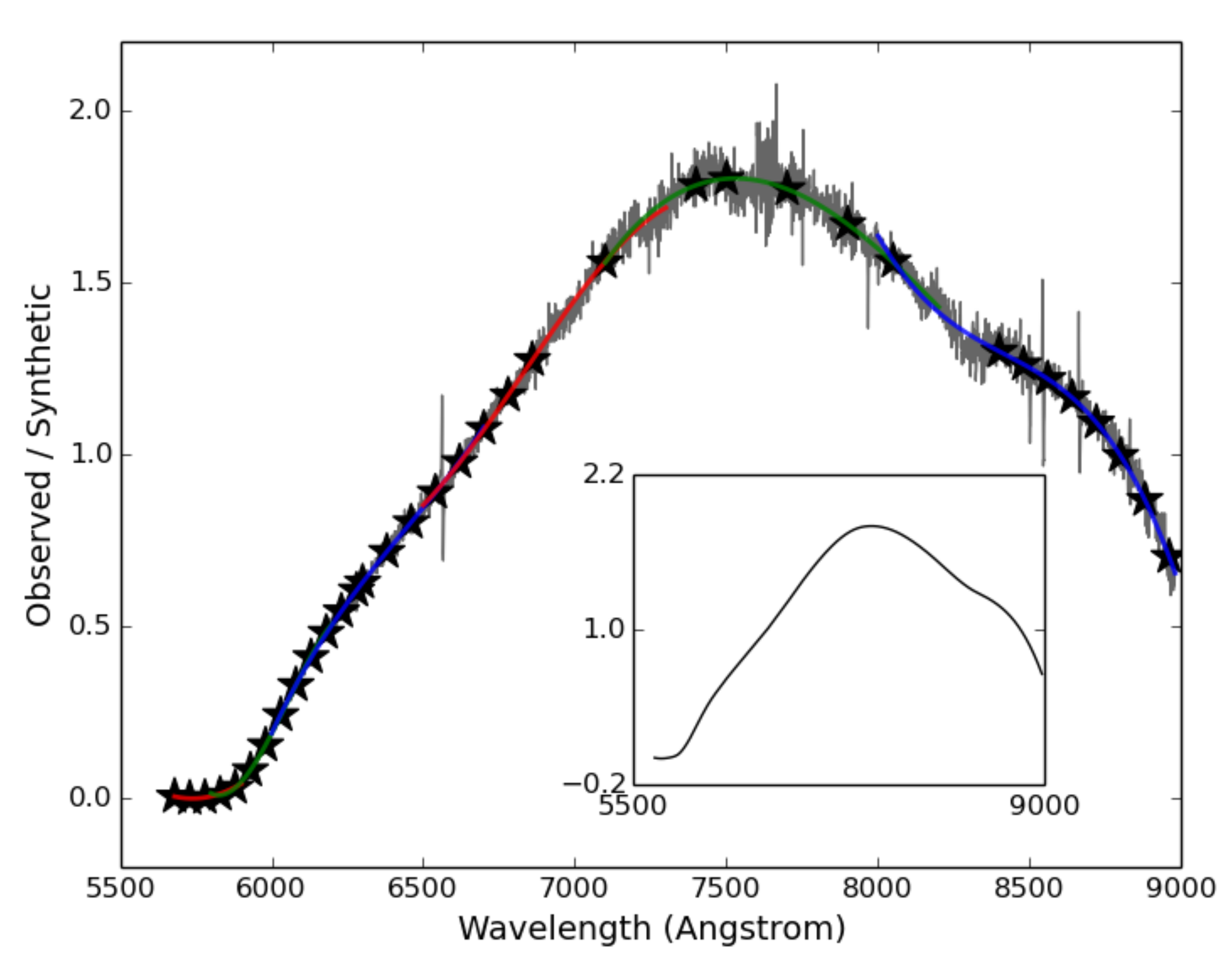}
\caption{Examples of SRC fitting for blue-- (left) and red--arm (right) spectra. The grey lines are the ratios of observed flux density and the synthetic flux density, the ratios have been scaled by their mean value. The blue-- and red--arm spectra are divided into five and six wavelength regions respectively, and each region is fitted with a second-- or third--order polynomial, which are represented by the thick lines in RGB colors (e.g. The blue arm has 5 bins drawn in the order of red, green, blue, red and green; while the red arm has six bins drawn in the order of red, green , blue, red, green, and blue).  Asterisks are selected from the fitted polynomial curves avoiding prominent spectral features, which are used for the final SRC fitting.}
\label{fig3}
\end{figure}

It is generally assumed that the SRCs are smooth functions of wavelength. In order to derive the SRCs, we applied a low-order piecewise polynomial fitting to the ratios of the observed and the synthetic spectra of the standard stars. Fig \ref{fig3} shows examples of the SRC fitting for one fiber in both arms. For each standard star, the blue-- and red--arm spectra were  divided into five and six wavelength regions respectively, and each region was fitted with a second-- or third--order polynomial, which are represented by the  thick colored lines in Fig \ref{fig3}. The piecewise polynomials were derived through minimizing $| synthetic\times polynomial-observed |$. We defined a series of clean spectral regions avoiding the prominent stellar absorption features and the telluric absorption bands. The fitted polynomial values of data in these clean regions is indicated by asterisks in Fig \ref{fig3}, which were used for the final SRC fitting. The wavelengths of the join points were fixed to space between 200 \AA\  for adjacent spectral regions,  the overlaps were median filtered  to join together the adjacent regions.  The final SRCs are represented by the black curves in the inserts in Fig \ref{fig3}.

\subsection{Derivation of the ASPSRCs}

\begin{figure}
%\begin{tabular}{cc}
\begin{minipage}{0.3\linewidth}
  \centerline{\includegraphics[width=5.5cm]{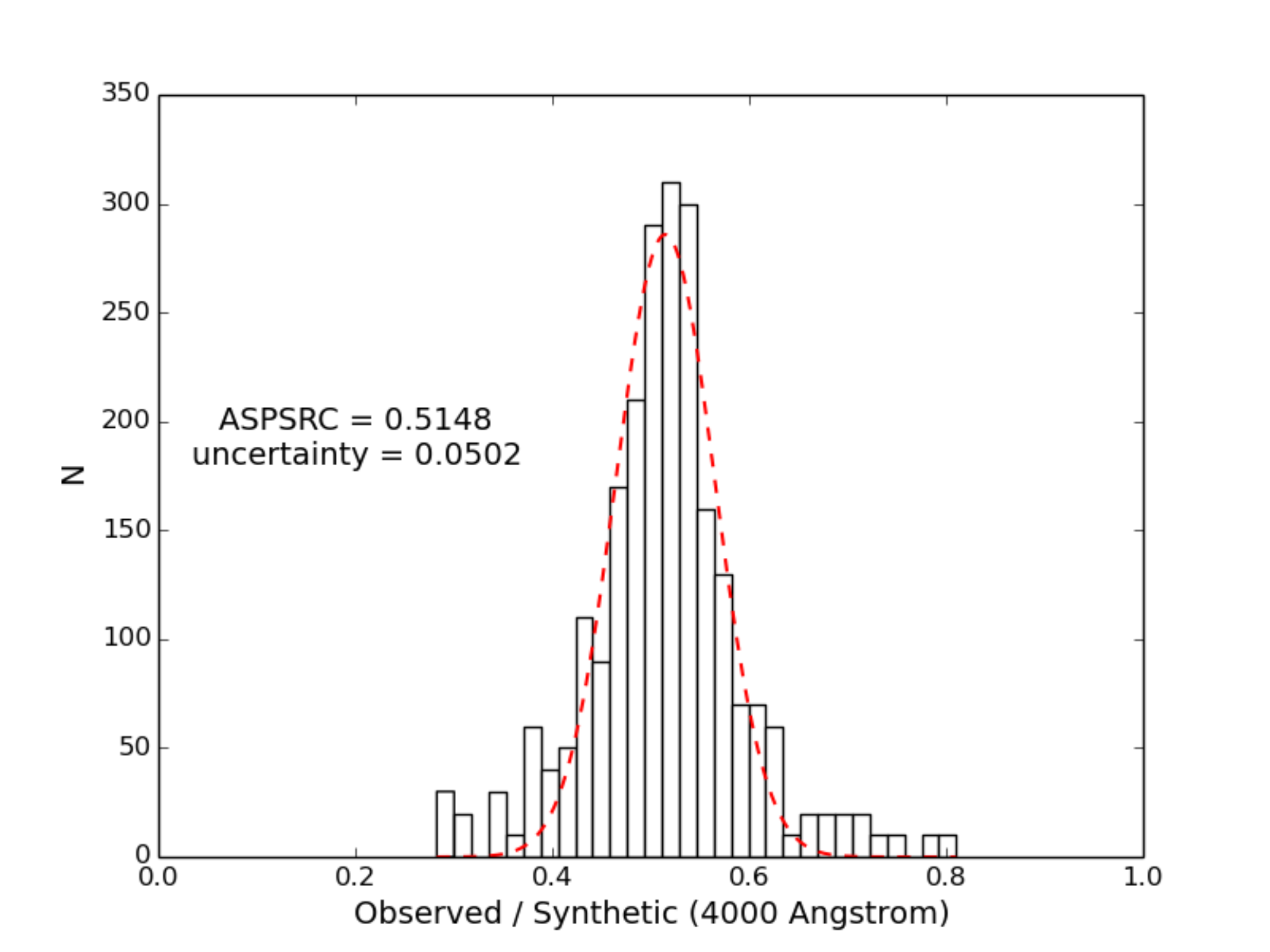}}
  %\centerline{(a) J064815.90+080240.7}
\end{minipage}
\hfill
\begin{minipage}{0.3\linewidth}
  \centerline{\includegraphics[width=5.5cm]{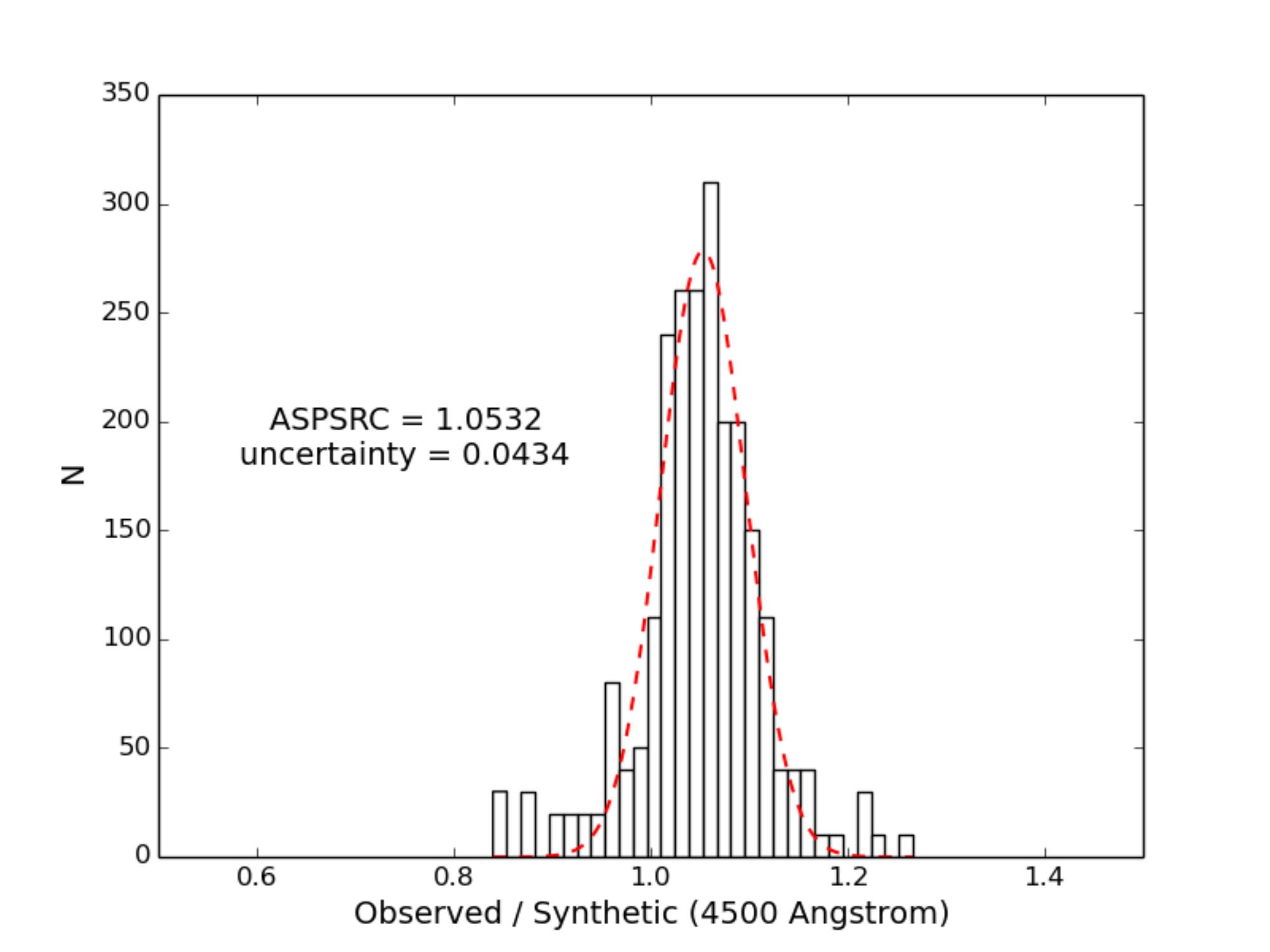}}
  %\centerline{(b) J040401.78+271545.4}
\end{minipage}
\hfill
\begin{minipage}{0.3\linewidth}
  \centerline{\includegraphics[width=5.5cm]{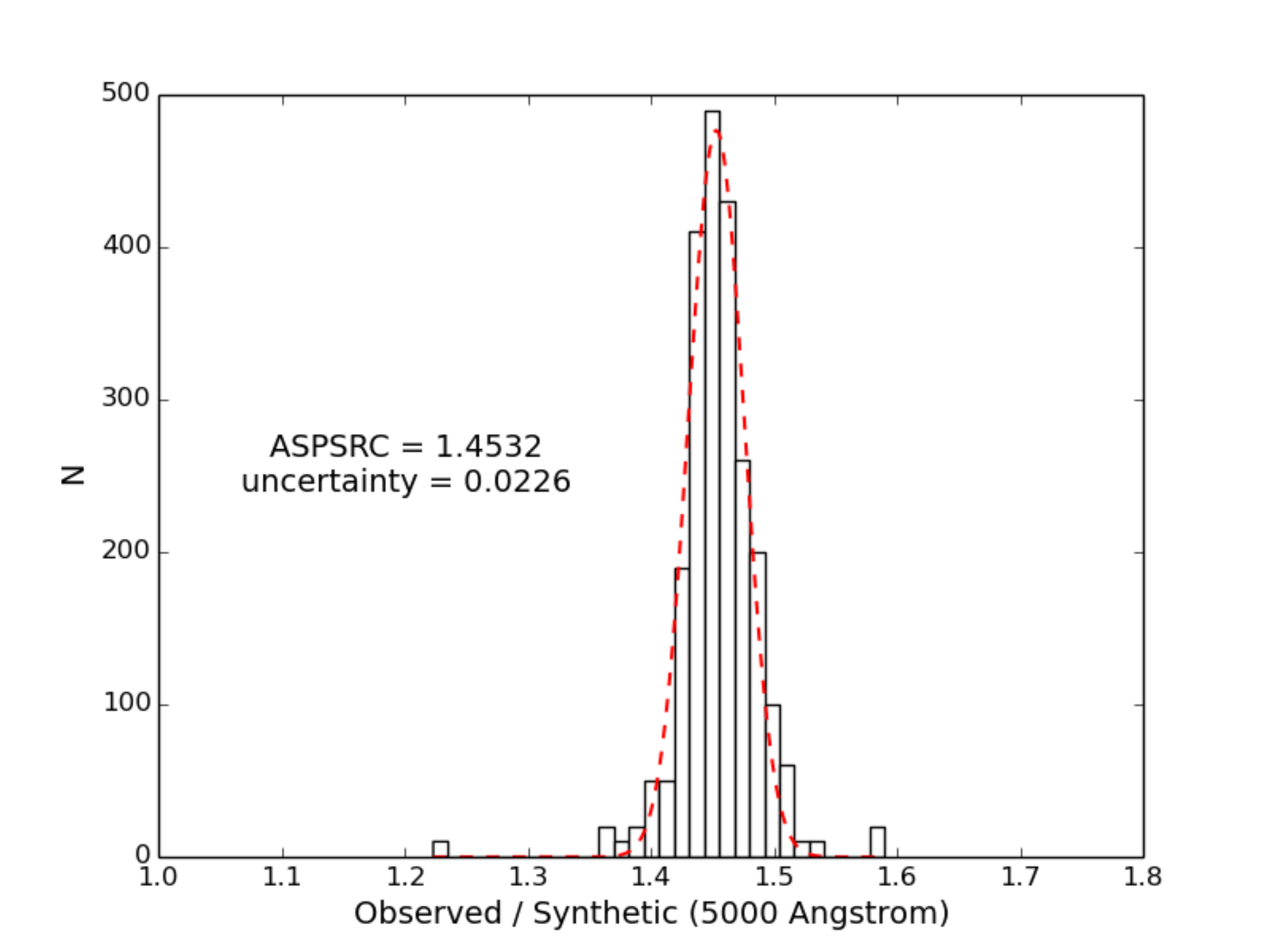}}
  %\centerline{(b) J040401.78+271545.4}
\end{minipage}
\label{fig4}
\caption{Histograms of SRCs  in 4,000 \AA (left), 4,500 \AA (middle) and 5,000 \AA (right) of  spectrograph No.1, the red dashed curves are Gaussian fits to the distributions, the mean and dispersion of the Gaussian fit to the ASPSRC and uncertainty are also marked.}
\end{figure}

\begin{figure}
\centering
\includegraphics[width=10.0cm]{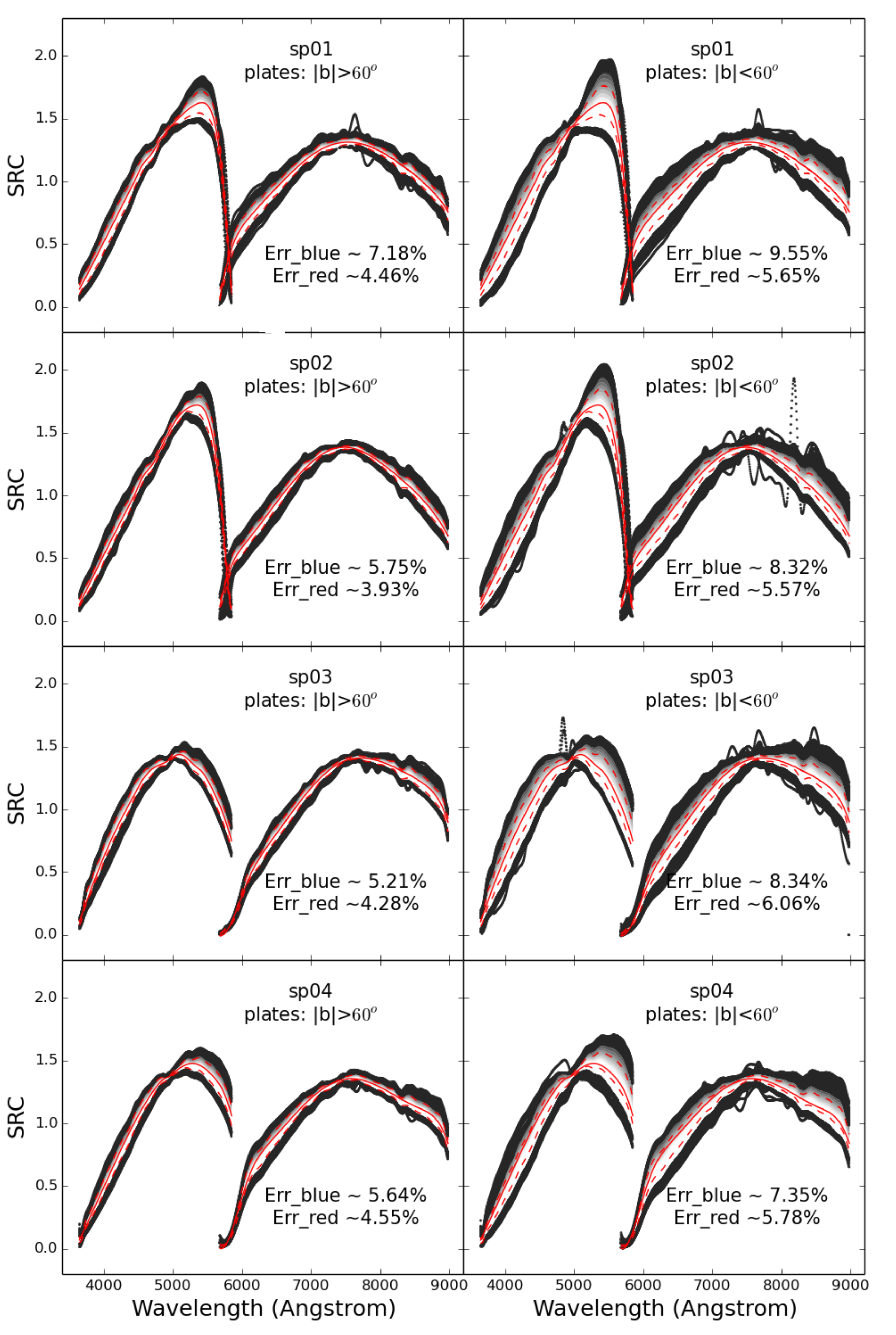}
\caption{The distributions of SRCs of spectrographs from   No.1 to No.4, the SRCs were derived from 2D pipeline for the DR2 plates. The grey contours represent the distributions of SRCs from 372 plates with 914 exposures for high galactic latitude  in left panel,  and  the distributions of SRCs from 1,759 plates with 3,387 exposures for the low galactic latitude  in right panel. The standard deviation of the SRCs as a function of wavelength is shown by the dashed curves, and the ASPSRCs we described in this paper  are  shown by solid curves.}
\label{fig5}
\end{figure}

\begin{figure}
\centering
\includegraphics[width=11.0cm]{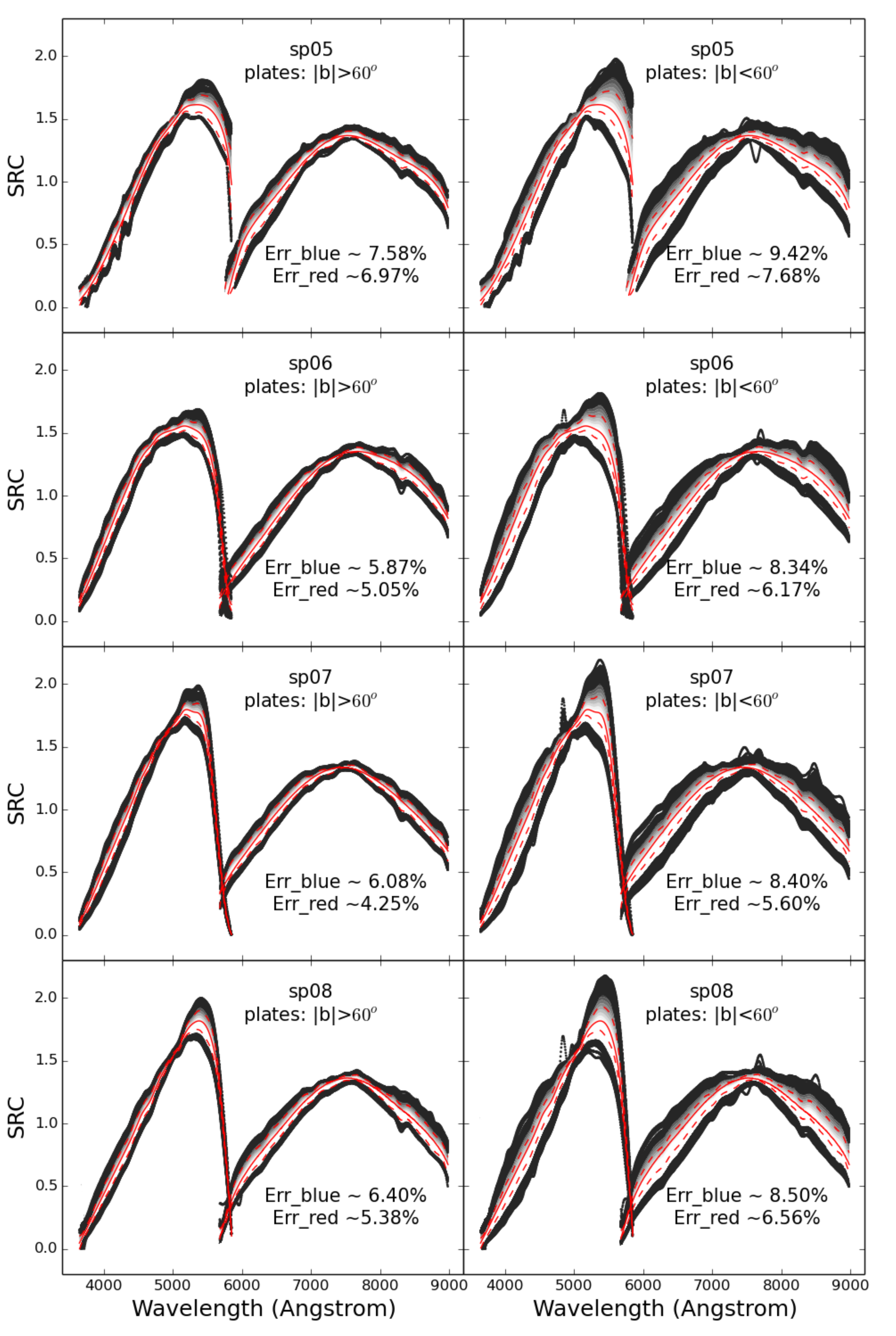}
\caption{The distributions of SRCs of spectrographs from  No.5 to  No.8. The convention is the same as in Fig \ref{fig5}. }
\label{fig6}
\end{figure}

\begin{figure}
\centering
\includegraphics[width=11.0cm]{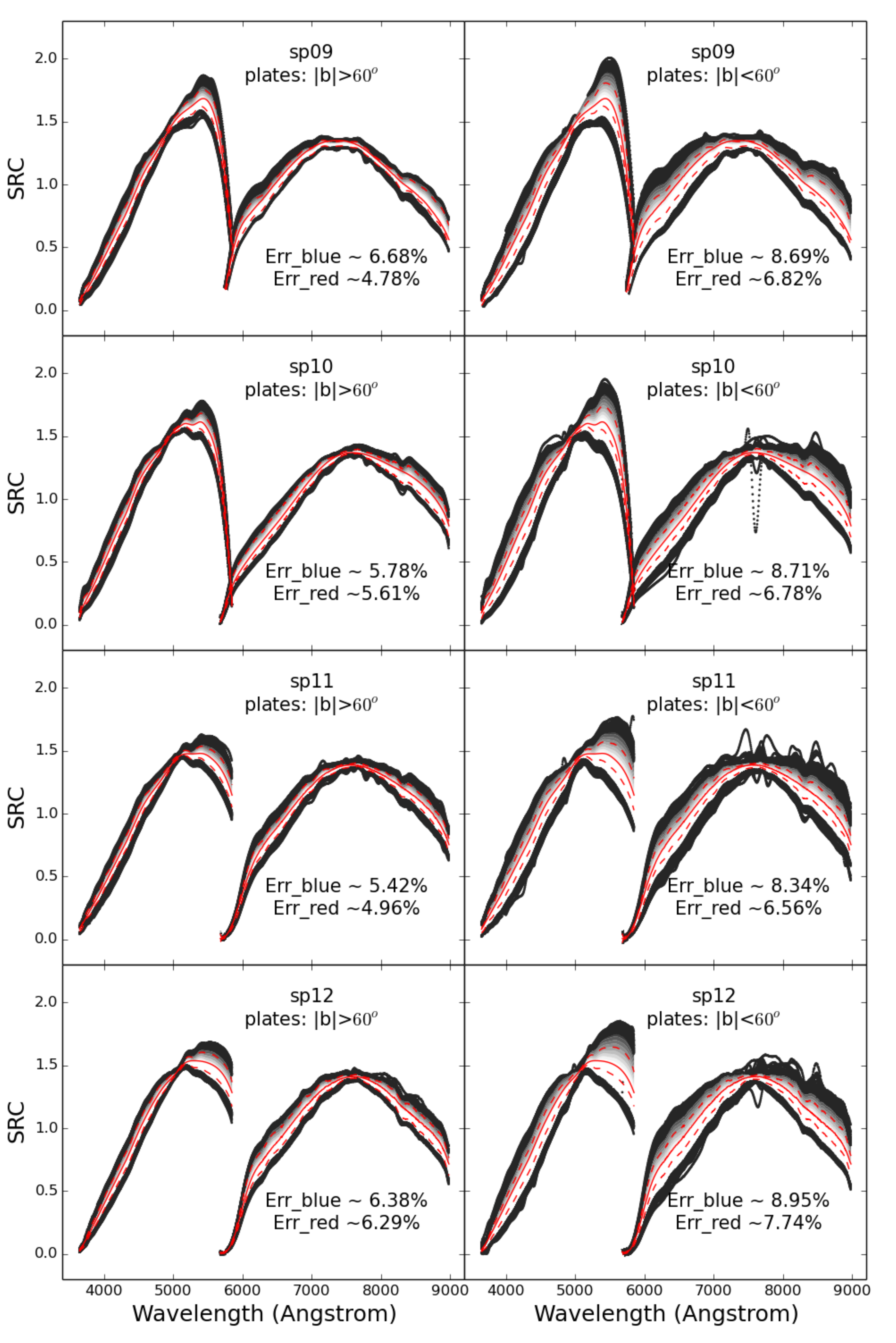}
\caption{The distributions of SRCs of spectrographs from  No.9 to  No.12. The convention is the same as in Fig \ref{fig5}.}
\label{fig7}
\end{figure}

\begin{figure}
\centering
\includegraphics[width=11.0cm]{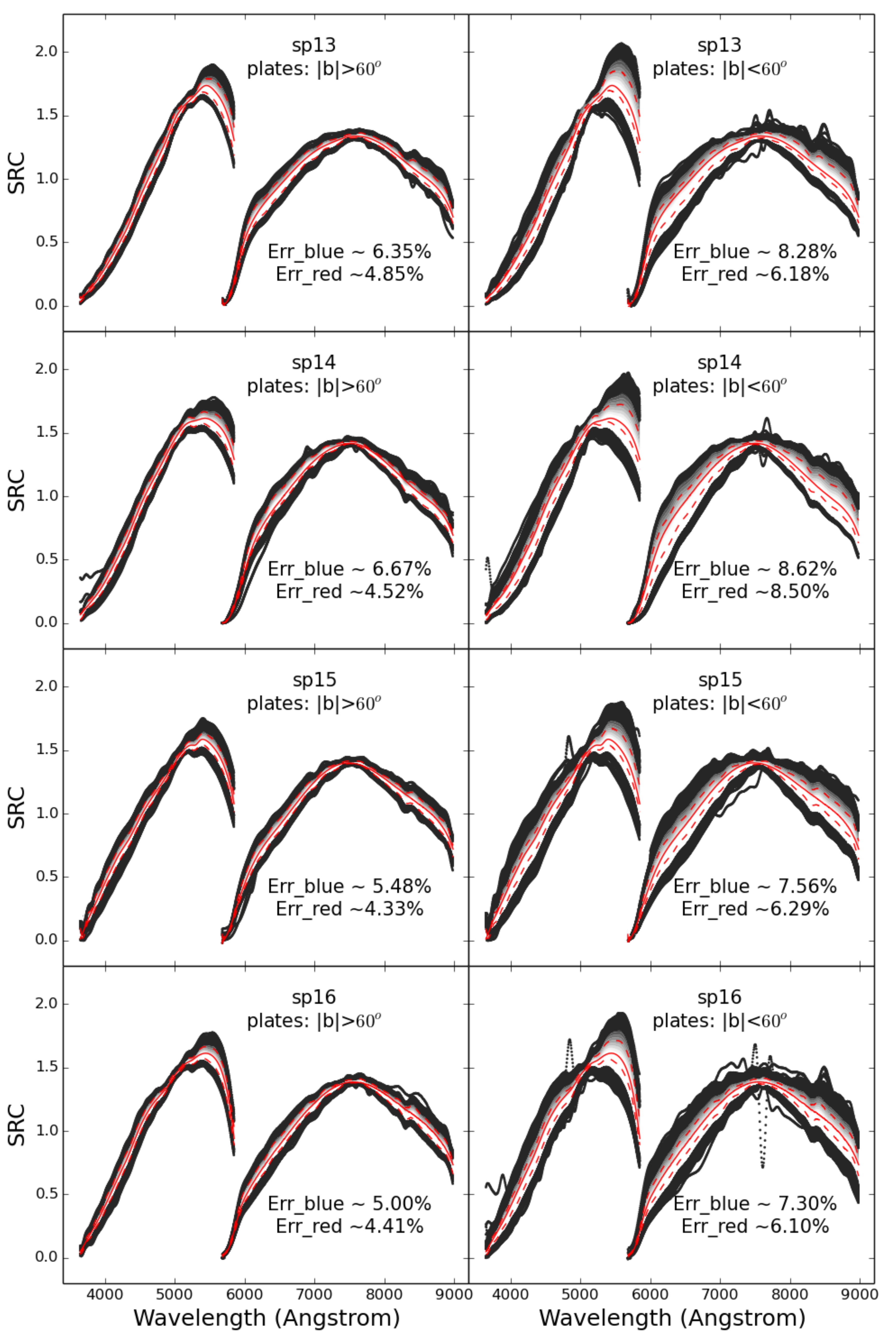}
\caption{The distributions of SRCs of spectrographs from  No.13 to  No.16. The convention is the same as in Fig \ref{fig5}.}
\label{fig8}
\end{figure}

For the 250 fibers in each spectrograph, at least 1800 good SRCs (through multi--exposures) were derived (excluding the bad fibers).  We chose the fitted SRCs rather than the direct ratios of observed and synthetic to estimate the ASPSRC  because the direct ratios are susceptible to noise.  We concentrate on the relative flux calibration rather than absolute flux calibration, such that,  for a given spectrograph, SRCs yielded by the spectra of the individual standard stars were divided by the average of their SRCs (i.e. the SRCs were scaled to a mean value of unity).  It is generally assumed that the differences in the sensitivity of the individual fibers are well corrected via flat--fielding and thus the 250 fibers of a given spectrograph share a single SRC.  Accordingly, the SRCs of the fibers can be regarded as independent measurements of the SRC, thus the ASPSRC and uncertainties can be derived by traditional statistical methods. The means and standard deviations of the Gaussian functions in Fig \ref{fig4} give three examples of the spectral response and uncertainty estimation at wavelength point 4,000 \AA, 4,500 \AA, and 5,000 \AA \ of  spectrograph No.1. All wavelength points contribute to the final ASPSRC for a spectrograph, and  the red curves in Fig \ref{fig5}, Fig \ref{fig6}, Fig \ref{fig7} and Fig \ref{fig8} show the blue and red--arm ASPSRCs of the 16 spectrographs.

We expect that there is a unified response for a given spectrograph during exposures of different times and using different plates. The red curves in Fig \ref{fig5}, Fig \ref{fig6}, Fig \ref{fig7} and Fig \ref{fig8} are more likely to be, at least, very close to the SRCs of the instrument including the atmospheric extinction, which means that the ASPSRCs derived can be used to calibrate the plates which lack  standard stars.  Some ASPSRCs drop deeply on the edges due to the very low sensitivities of the instrument, which is the most challenging problem for LAMOST flux calibration.

We calculated the mean values of absolute and relative uncertainties for  g, r and i--bands, which are presented in Table \ref{tab1}.  Table \ref{tab1}
shows that for all 16 spectrographs, the uncertainties are smaller than 8\% for both g and i--bands. The r--band is
located at the edge of both arms and thus due to the low sensitivities, the uncertainties for r--band are much larger ( for example Spectrograph No.5 can differ by
up to 11.13\%.  This means the fluxes and centroids of the lines located at the junction of the blue and red arms (such as Na D at $\lambda$5,892\AA), are sometimes not credible.

\subsection{Time Variations}
\begin{figure}
\centering
\includegraphics[width=11.0cm]{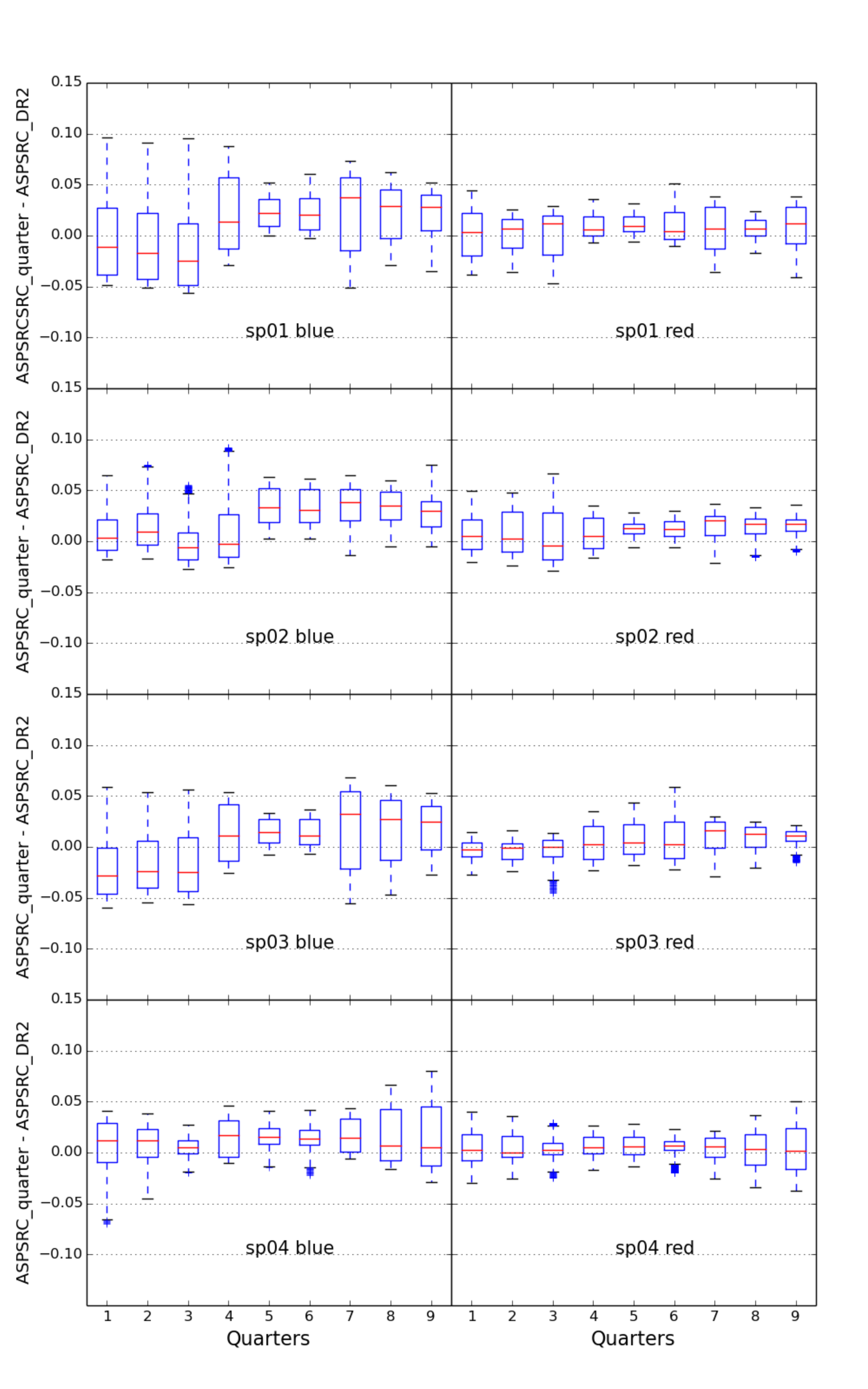}
\caption{The distributions of residuals between the nine Quarter ASPSRCs and the DR2 ASPSRC (blue arm in left panel and red arm in right panel), for spectrographs from No.1 to No.4. The box extends from the lower to upper quartile values of the error, with a line at the median. The whiskers extend from the box to show the range of the error. Flier points are those past the end of the whiskers.}
\label{fig9}
\end{figure}

\begin{figure}
\centering
\includegraphics[width=11.0cm]{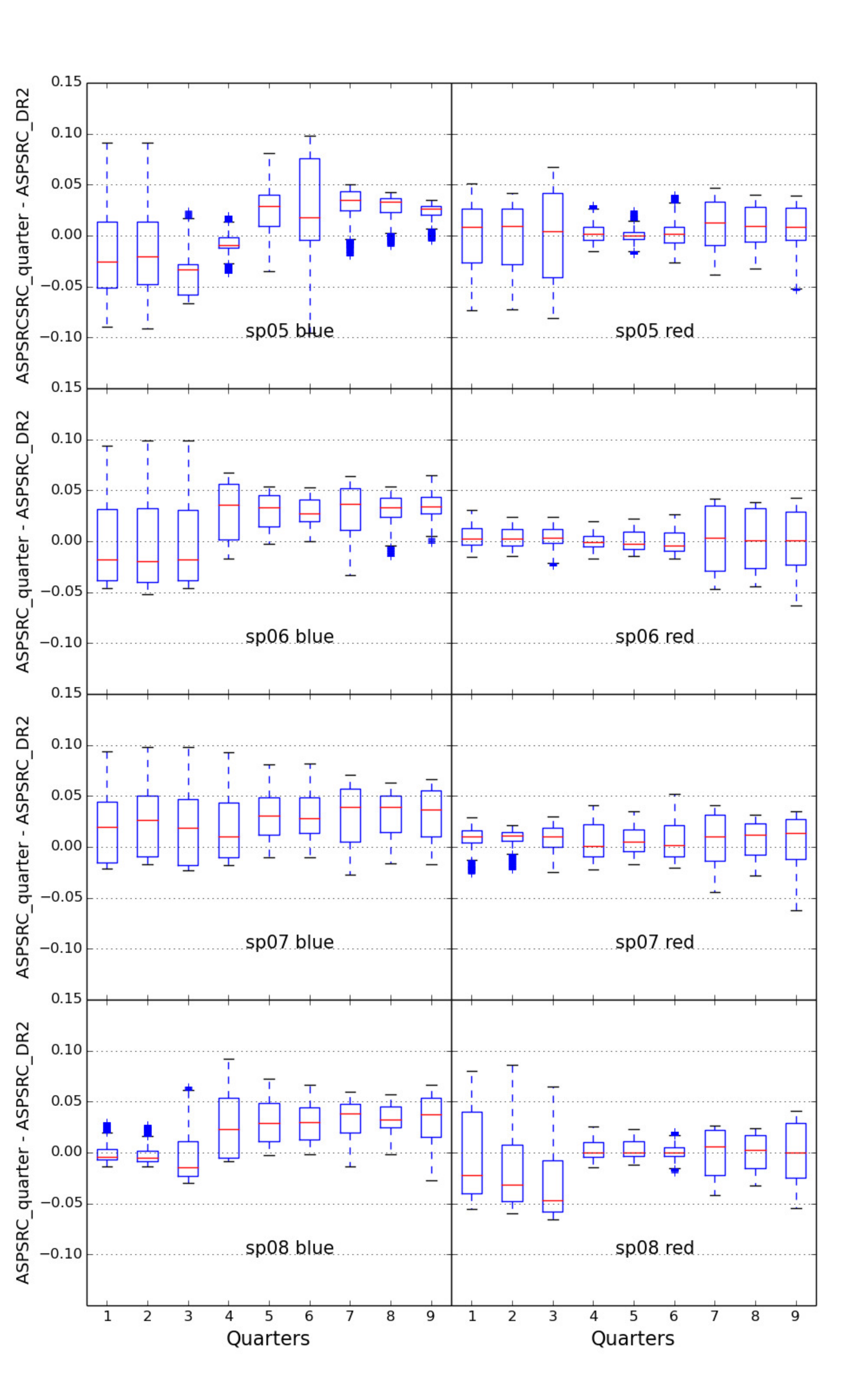}
\caption{The distributions of errors for spectrographs from  No.5 to  No.8. The convention is the same as in Fig \ref{fig9}. }
\label{fig10}
\end{figure}

\begin{figure}
\centering
\includegraphics[width=11.0cm]{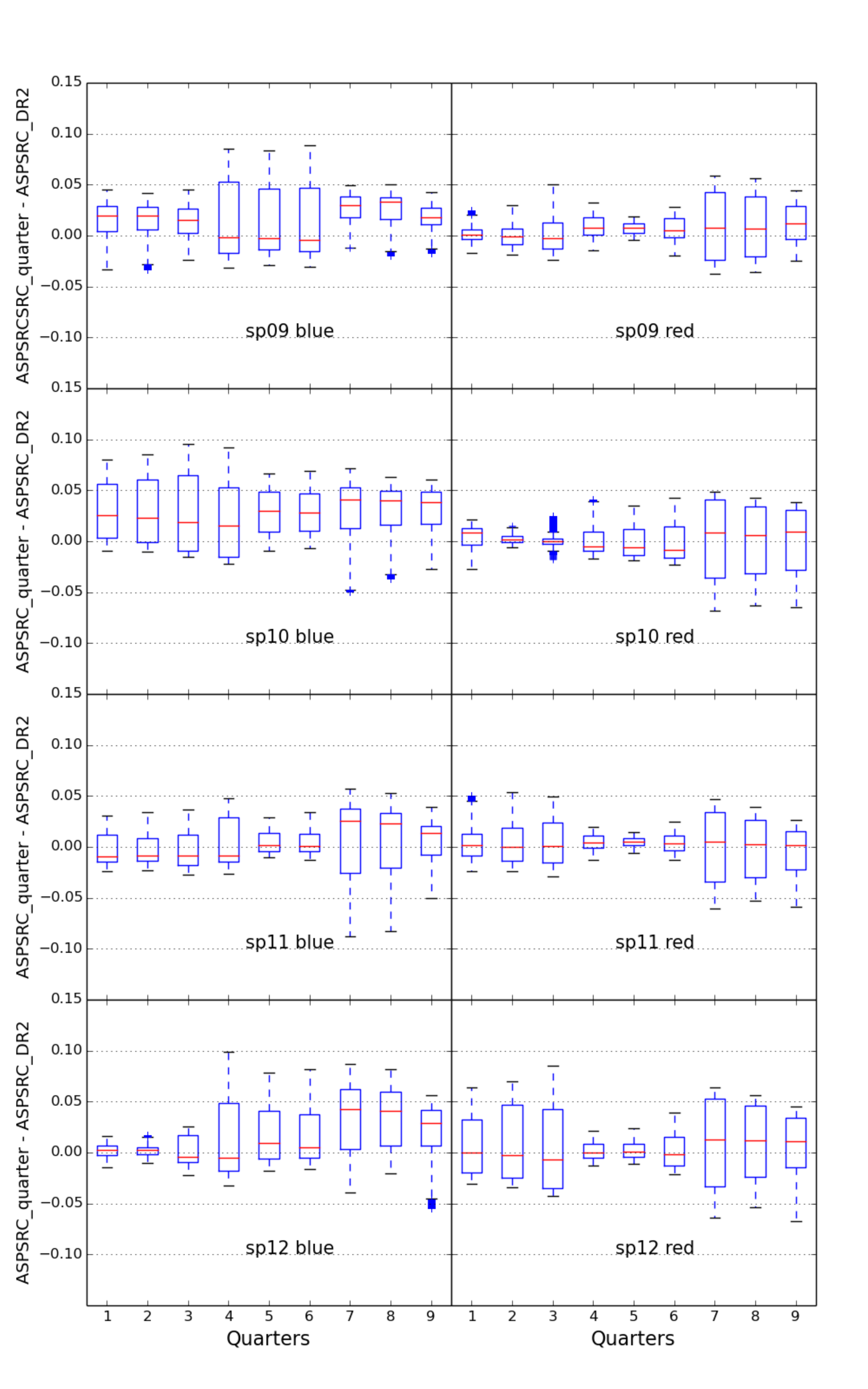}
\caption{The distributions of errors for  spectrographs from  No.9 to  No.12. The convention is the same as in Fig \ref{fig9}.}
\label{fig11}
\end{figure}

\begin{figure}
\centering
\includegraphics[width=11.0cm]{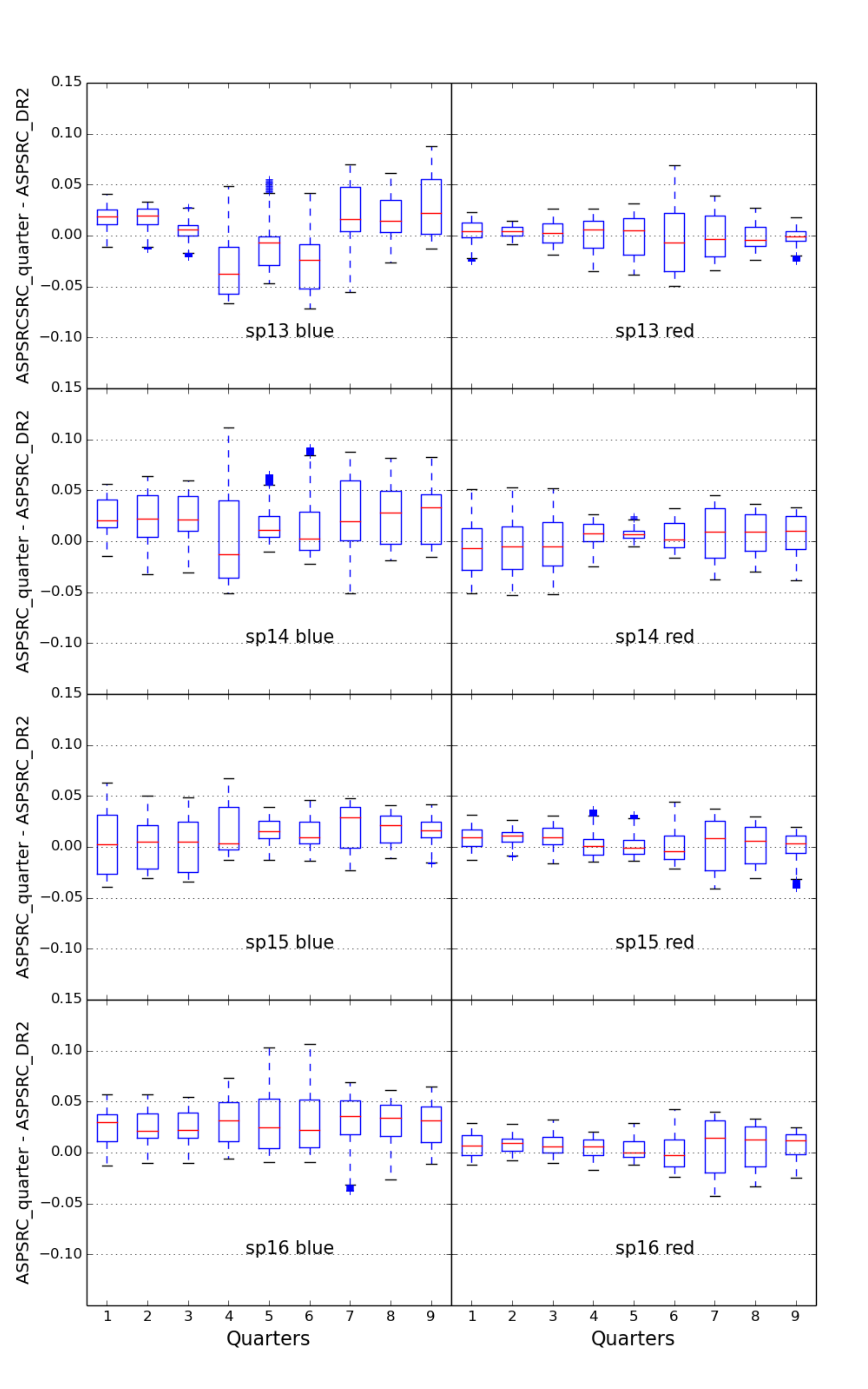}
\caption{The distributions of errors for spectrographs from  No.13 to  No.16. The convention is the same as in Fig \ref{fig9}. }
\label{fig12}
\end{figure}

Generally, the LAMOST observational season spans  nine months from September to next June. The DR2 collected the  observed data from October 2011 to June 2014, and there are nine quarters totally (about 3 consecutive months for a quarter).  We calculated ASPSRCs for each quarter  (hereafter called Quarter ASPSRC to distinguish from DR2 ASPSRC), and  compared these nine Quarter ASPSRCs with the DR2 ASPSRC for each spectrograph. The distributions of residual between the nine Quarter ASPSRCs and the DR2 ASPSRC are shown in Fig \ref{fig9}, Fig \ref{fig10}, Fig \ref{fig11} and Fig \ref{fig12} (blue arm in left panel and red arm in right panel).  The figures show that there are not obvious gradual and systematic errors along with time. Still, we can conclude that spectrograph No.4, No.11, No.15 and No.16 are more stable than others during the DR2 period. 

\section{Flux Calibration Based on ASPSRCs}
\label{sect:calibration}
The spectral flux calibration of target objects is generally achieved through obtaining separate measurements of spectrophotometric standard stars \citep{Oke,H1,H2} on the same observing night with the same instrumental setup.  However large spectroscopic survey, obtaining separate measurements of sufficient standard stars for each night and each spectrograph becomes impossible, and an alternative strategy has to be adopted. In the case of the Sloan Digital Sky Survey \citep{York}, F turn--off stars within the  FoV are used to calibrate the spectra. These standards are preselected based on the photometric colors and are observed simultaneously with the targets  \citep{Stoughton,Yanny}. The intrinsic SEDs of F turn--off stars are well determined by theoretical models of stellar atmospheres and the effects of interstellar extinction can be characterized and removed using the all-sky extinction map of Schlegel  et al. (1998) \citep{schlegel,schlafly}. Without a photometric system for LAMOST, and lacking of extinction values especially for low galactic latitudes, the standard stars are not pre--assigned. Usually, the flux standard stars are selected from the spectra in each spectrograph after observation. Sometimes, the selection of the standard stars fails,  thus the spectrograph of  plate has to be abandoned by the LAMOST 2D pipeline. This is indeed why the ASPSRC method is important, as using fixed instrumental response curves can recover some of these abandoned plates.

\subsection{Co--add the multi--exposures}

\begin{figure}
\centering
\includegraphics[width=12.0cm]{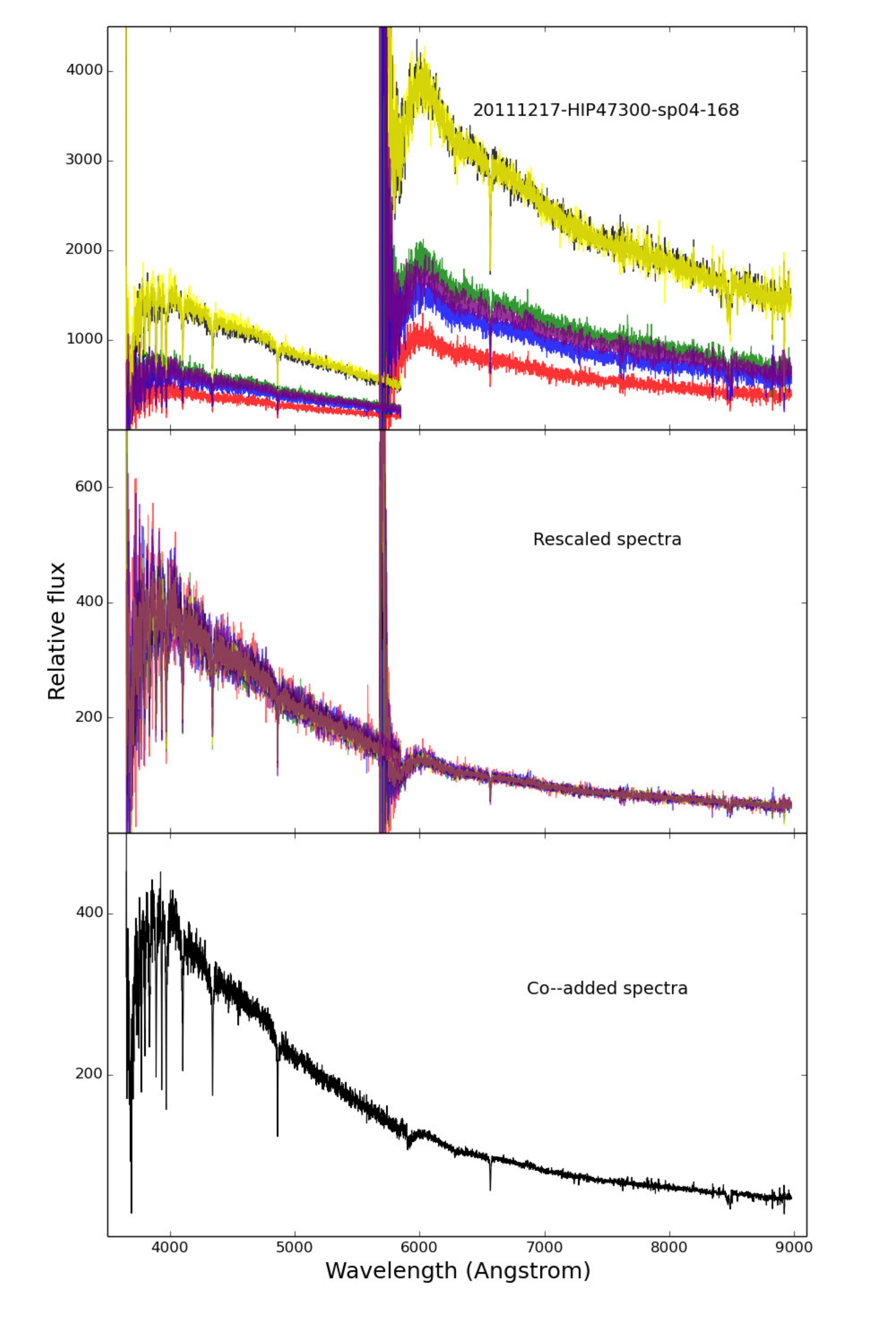}
\caption{Example of a star with 6 exposures in different scales, for blue and red--arms, the spectra of  equal exposure are plotted by the same color (top). The rescaled spectra (middle) are scaled depend on g and i magnitudes. The co--added spectra (bottom) are adopted as the final spectra. }
\label{fig13}
\end{figure}

To improve the SNRs and overcome the effect of cosmic rays, each field is designed to be exposed multiple times. The spectra of each single exposure may be on different scales due to the variation of observational condition. The spectra on different scales can not be co--added since they are divided by the same ASPSRC. Fig \ref{fig13} shows the spectra of 6 exposures. Not only are the scales of exposures  different but also the scales of the two arms are discrepant.  For the LAMOST 2D pipeline, the single exposure spectra are scaled to the median of the multi--exposures. Here we try to scale the blue and red band according to the photometry of the g and i band respectively.

 The monochromatic AB magnitude is defined as the logarithm of a spectral flux density with a zero-point of 3631 Jansky \citep{oke83}, where 1 Jansky = 1 Jy = $10^{-26} W Hz^{-1} m^{-2} = 10^{-23} erg s^{-1} Hz^{-1} cm^{-2}$.  If the spectral flux density is denoted $f_\nu$, the monochromatic AB magnitude is:
\begin{equation}
   m_{AB}(\nu) = -2.5 \ log_{10}f_\nu - 48.60
\label{eq:eq4}
\end{equation}

Actual measurements are always made across some continuous range of wavelengths. The bandpass AB magnitude is defined similarly, and the zero point corresponds to a bandpass--averaged spectral flux density of 3631 Jansky.

\begin{equation}
   m_{AB} = -2.5 \ log_{10}(\frac{\int f_\nu(h\nu)^{-1}e(\nu)d\nu}{\int 3631Jy (h\nu)^{-1}e(\nu) d\nu })
\label{eq:eq5}
\end{equation}

where $e(\nu)$ is the equal--energy filter response function. The $(h\nu)^{-1}$ term assumes that the detector is a  photon-counting device such as a CCD or photomultiplier. The synthetic magnitude can be obtained by convolving the flux spectra with the SDSS g and i band transmission curves \citep{H1,H2}. We adopted the g and i filter zero points from Pickles \citep{pickles}.  The spectra  are then scaled by comparing the synthetic magnitude with the photometry magnitude. The scale coefficient $SC(g)$ and $SC(i)$  are obtained as follows, and are multiplied with observed spectra.
\begin{equation}
   SC(g) = 10^{-0.4\times[mag(g)-mag_{synthetic}(g)]}; \   \
   SC(i) = 10^{-0.4\times[mag(i)-mag_{synthetic}(i)]}
\label{eq:eq6}
\end{equation}

 The spectra of  Fig \ref{fig13} were scaled using the method described above. The rescaled spectra  can then be co--added, and the final spectra derived, which is shown in Fig \ref{fig13} (bottom panel). It should be noted that this method is subject to the SNR of the spectra, since the synthetic magnitudes depend on the quality of the spectra.

This method needs the photometry magnitudes of g and i--band for each target, thus we cross--matched LAMOST targets with Pan--STARRS1 \citep{tonry} within 3 mas. The LAMOST sources are selected from multiple catalogs with multi--band photometry. Consequently, not all the LAMOST targets overlap Pan--STARRS1. By cross--matching, we found that about 80\% of the LAMOST targets are included in Pan--STARRS1. For those targets not in the Pan--STARRS1 catalog, SDSS PetroMag ugriz had been adopted. About 60\% of the LAMOST targets have the overlapping observations with SDSS. However, there are still about 10\% of the LAMOST targerts neither included in Pan--STARRS1 nor in SDSS, and it is difficult to obtain their photometry in the optical band. For the LAMOST, the efficiencies of the blue and red--arms can not be corrected by flat fielding since the throughput of the two arms for each spectrograph are different and varies when the telescope pointing changes.  The flat field of the two arms for each spectrograph are processed independently. Using photometry, we can avoid big scale jump between the two arms although there are some photometry errors. Without the reference of photometry, we have to only use the overlap between the blue and red arms in a very small wavelength range to connect them, that might lead to an piecing discontinuity if the signal noise ratio is too low in the overlap.

For the final spectra, spline fitting with strict flux conservation is adopted to re--bin the spectra to a common wavelength grid. Once the flux is co--added by this method, the blue-- and red--arms are pieced together directly and the SEDs are consistent with their target colors. For the ones which don't have photometry in the optical band, but have multiple exposures, we scaled the flux of multi--exposures to the flux of exposure with the highest SNR. After the multi--exposures being co--added, The blue and red-arm are pieced together with adjusting one of the scale (using the overlaps) to yield the final spectra.

\subsection{Accuracy analysis for flux calibration through ASPSRC }
Before discussing the accuracy of the ASPSRCs, we studied the SRCs of the DR2 plates which are derived from the LAMOST 2D pipeline to further confirm  the stability of the LAMOST spectrograph response curves.  Fig \ref{fig5} to Fig \ref{fig8} shows the distributions of the SRCs of the DR2 plates for high Galactic latitude (left panel) and for low Galactic latitude (right panel), the standard deviations of the SRCs,  as a function of wavelength,  are shown by the red dashed curves.

As described in Section 2.3, we used  stars with high Galactic latitude and with high SNR to get the ASPSRCs. The red solid curves in  Fig \ref{fig5} to Fig \ref{fig8} represent the ASPSRCs, and it is consistent with the average SRC of the SRCs from the LAMOST 2D pipeline.   Table \ref{tab1} shows  the mean uncertainties of ASPSRCs are smaller than 10\%,  which are consistent with the 1 $\sigma$  uncertainties of the SRCs at high Galactic latitude from the 2D pipeline.

\begin{figure}
\centering
\includegraphics[width=16.5cm,height=9.0cm]{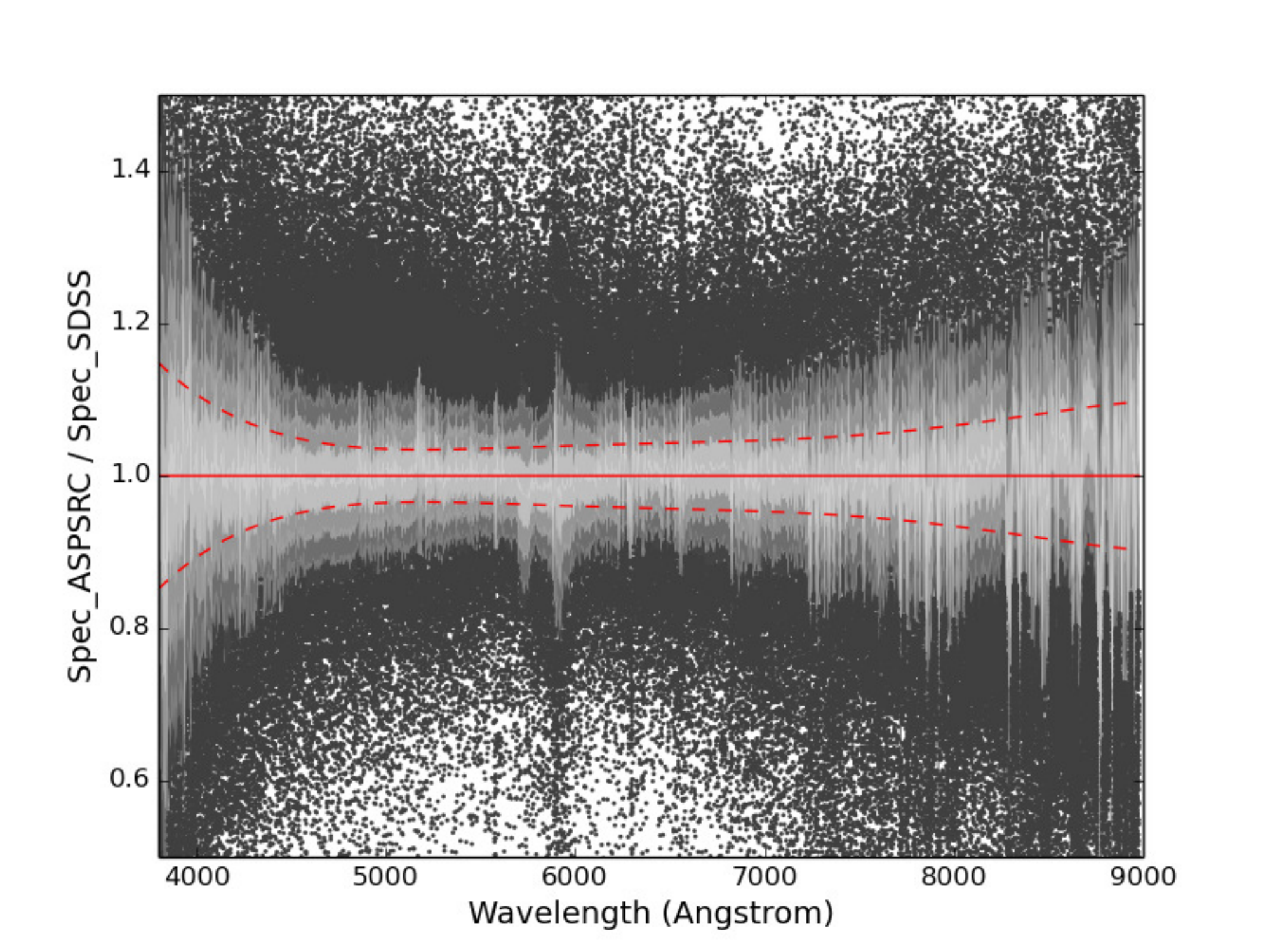}
\caption{Distribution of  the ratios of spectral pairs observed by both LAMOST and SDSS, each point on the panel is a ratio value of one pair, the contour represent the distribution of the ratio values of 1,746 spectral pairs. The smoothed mean and standard deviation of the ratios,  as a function of wavelength,  are shown by the solid and dashed curves.}
\label{fig14}
\end{figure}

To verify the feasibility of applying the ASPSRCs to the flux calibration, we selected stars observed by both LAMOST and SDSS. We cross--matched the abandoned targets of LAMOST DR2 with SDSS DR12,  and obtained 1,746 spectra of 1,702 stars with SNRs higher than 6. We have calibrated the LAMOST spectra abandoned by 2D pipeline and divided them by the spectra of the same sources from SDSS. The ratios of the two spectra were calculated and then scaled to their median values of unity, and the results are shown in Fig \ref{fig14}. The ratios yield an average that is almost constant around 1.0 for the whole spectral wavelength coverage except for sky emission lines region; oxygen and water vapor bands of the earth's atmosphere are attributed to the uncertainties of flat--fielding and sky--subtraction. The standard deviation is less than 10\%  at wavelengths from  4,500 \AA  \ to 8,000 \AA  \ but for both edges, the standard deviation increases to 15\% due to the rapid decline of the instrumental throughput. The results show that flux calibration using ASPSRCs  has achieved a precision of $\sim$ 10\% between 4,100 \AA  \ and 8,900 \AA.

\begin{figure}
\centering
\includegraphics[width=18.0cm,height=18.0cm]{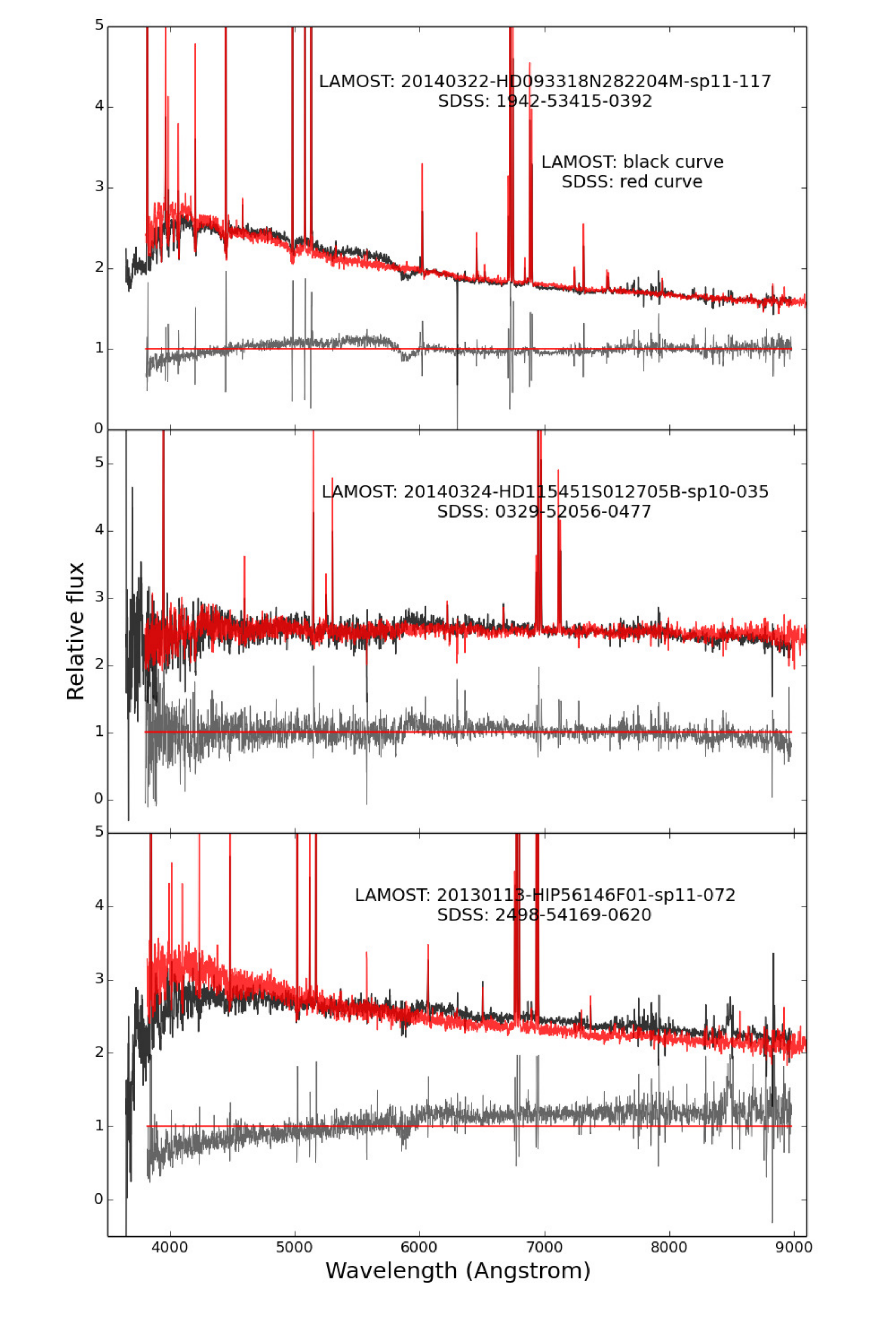}
\caption{Comparison of the rescued spectra of galaxies (black)  with SDSS DR12 spectra (red). For each panel, the upper part shows the relative flux density as a function of wavelength, whereas the lower part shows the ratios of LAMOST and SDSS.}
\label{fig15}
\end{figure}

For the bright and very bright plates, most can be calibrated successfully by 2D pipeline. However for LAMOST faint plates (F--plates) of DR2, the flux--calibration failure rate of the 2D pipeline is around 9\%£¬ and for the medium plates (M--plates), the failure rate is around 8\%.  Fig \ref{fig15} to Fig \ref{fig17} show the spectra of galaxies, QSOs and stars rescued from the abandoned plates. We compared the rescued spectra with that of SDSS DR12 (the former are plotted with black curves and the latter are represented with red curves). Most match with their corresponding SDSS spectra quite well, with differences of only a few percent for their continua. For LAMOST 20130208-GAC062N26B1-sp13-112, the spectra of the red arm  has turbulent components for the spectrograph No.13, this is explained by the spectrograph having problems caused by the cooling system of the CCD. For  LAMOST 20140306-HD134348N172427B01-sp10-014, the SED from the ASPSRC method is bluer than that of SDSS.  We believe  this is due to the fact, we do not separate the Earth's atmospheric extinction from the response of  the spectrograph. Generally, the variations of the optical atmospheric extinction curve can be  calculated by low order polynomials \citep{patat}. The atmospheric extinction curve included in the ASPSRC is an average one, and multiplication by a low order polynomial is required to obtain the real atmospheric extinction curve when the target observed. Therefore, some spectra calibrated using the ASPSRCs need low order polynomials to match SDSS spectra. The atmospheric extinction of LAMOST will be deeply studied and integrated into this work.

Overall, the ASPSRCs flux calibration has achieved a precision of $\sim$ 10\% for the LAMOST wavelength range. The potential uncertainties and temporal variations of the atmospheric extinction generally do not have an impact on the final accuracy of spectral lines, though they do affect the shapes of SEDs deduced (low order polynomials).

\subsection{Rescue the Abandoned Targets}

For  the LAMOST DR2, there are 1,095 spectrographs with 385 plates which have been abandoned by 2D pipeline due to the failure of finding standard stars.  We started with the 2D pipeline for fiber tracing, flux extraction, wavelength calibration, flat fielding and sky subtraction.  The ASPSRCs were then adopted to calibrate the 195,694 spectra in 1,095 spectrographs. After the flux calibration and the co--add, the LAMOST 1D pipeline was employed to classify the spectra and measure the radial velocity for stars,  and the redshift for galaxies and QSOs. Based on a cross--correlation method, the 1D pipeline recognizes the spectral classes and simultaneously determines the radial velocities or redshifts from the best fit correlation function. The 1D pipeline produces four primary classifications, namely STAR, GALAXY, QSO and UNKNOWN.

It is difficult to recognize galaxy and QSO spectra and determine their redshift, and as such the LAMOST 1D pipeline does not work  as well as  for stellar classification due to the SNRs of  galaxy and QSO spectra being relatively lower. An additional independent pipeline, the Galaxy Recognition Module (GM for short), has been designed for galaxy recognition and redshift measurement.  After the 1D pipeline was run it automatically identifies galaxies and measures the redshifts  by recognizing lines. The redshifts of galaxies are determined through line centers. Before line centers are measured, a Gaussian function with sigma of 1.5 times the wavelength step is applied to the spectra to eliminate noises. The continua, which were smoothed by a median filter, are divided to complete normalization. Those data points that exceed 2$\sigma$ of a normalized spectrum are selected as candidates of emission lines, then a set of Gaussian functions is used to fit the lines. All the line centers are compared with line lists, which are spaced by steps of 0.0005 in redshift($z$). If most of the lines are matched successfully with heavily weighted lines such as NaD, Mgb, CaII H or CaII K for absorption galaxies, or  H$\alpha$, OII, H$\beta$, OIII or NII for emission galaxies, the spectrum is classified as galaxy, and the corresponding $z$ is the raw redshift of the spectrum.  However, for QSOs, the classifications and measurements highly depend on visual inspection. We  combined the classification of GM, 1D pipeline and expert inspection and thus the final classifications of the spectra of the 1,095 spectrographs is presented in Table \ref{tab2}. In total 52,181 additional spectra has been recognized in DR2, and will be officially released in the Third Data Release (DR3) of the LAMOST Regular Survey. The fraction of objects rescued is about 52,000/2,000,000 ($\sim$ 2.5\%).

For the rescued 52,181 targets, we evaluated  the quality by plotting the magnitude against SNR relationships for galaxies and QSOs, and stars.
For galaxies and QSOs, most of the magnitudes are spread between 17.0 and 19.0, which is shown in Fig \ref{fig18}. This is close to the limit of LAMOST observation, consequently, the majority of their SNRs are so low that they do not reach a SNR of 10.  To reduce the differences of SNRs due to differences in exposure times, all of the SNRs in this paper were scaled to 5400s.  For stars, there are two peaks in the distribution of magnitudes, as shown in Fig \ref{fig19}. The magnitudes of A,F,G,K--type stars range from 13.0 to 17.0, and M--type stars range from 15.0 to 18.0 magnitudes. The SNRs of the stars are higher than that of galaxies and QSOs, however, most are below 30,  which is comparatively low for stars. With the exception of M--type stars, we selected the stars with SNRs in the r band larger than 2.0 for the release. Therefore, an obvious  cut is seen in the bi-modal point distributions of early and late type stars in Fig \ref{fig19}. For F,G,K--type stars, by running LASP, we parameterized those with SNRs in the g band larger than 6.0 for nights with a dark moon and 15.0 for  nights with a bright moon. The final stellar parameters coverage is presented in Fig \ref{fig20}.

\subsection{Revision of the 2D calibration}

To minimize potential errors introduced by poor sky subtraction, the current LAMOST 2D pipeline (v2.7) scales the sky spectrum to obtain the same flux intensities of sky emission lines as that of the target spectra, which it will be subtracted from. It is assumed that the emission lines are homogeneous across the FoV of individual spectrographs (about 1 deg). However, the continuum sky--background and sky emission lines originate from very different sources and are excited by different mechanisms, thus their emission levels are unlikely to scale linearly. In fact, even amongst the sky emission lines, lines from different species may have quite different behavior in terms of their temporal and spatial variations \citep{oliva}. Consequently, scaling the sky spectra by the measured fluxes of sky emission lines risks subtracting an incorrect level of sky background. For a minority of spectra the standard stars telluric bands are extremely under subtracted, and thus it turns out that the SRCs of the standards are over--fitted (see Fig \ref{fig21}). The oxygen band is under--subtracted for the spectra of the standards. This leads to the over fitted SRC also containing the oxygen band and introduces artificial spectral lines to all the spectra of the spectrograph, making the classification of spectra by 1D pipeline difficult.

Fig \ref{fig22} shows one example of an artificial spectra calibrated by the over--fitted SRC (from Fig \ref{fig21} using the 2D pipeline),  which is plotted with black curves. We recalibrated the spectra using the ASPSRCs,  which is presented with red curves in Fig \ref{fig22}. After recalibration, the spectra was classified as F0 by the 1D pipeline (an improvement over the `Unknown' classification by the 1D pipeline previously). Comparing the ASPSRCs with the SRCs from the LAMOST 2D pipeline, we found 6 spectrographs from DR2 have this problem, and all of the 6 plates were observed in the nights with very bright moon. The ASPSRC method has been used to correct this problem, and the spectra of this 6 spectrographs will be released in LAMOST DR3.

\section{Analysis and Discussions}
\label{sect:analysis}

We have applied the ASPSRCs to the flux calibration for LAMOST,  however,  there are still some uncertainties in the ASPSRCs caused by individual SRCs. The causes of these variations in the shape of SRCs might be attributed to several factors.

First of all, although we selected the standard stars from high galactic latitudes to minimize the affects from variations of interstellar extinction, the effect of the Earth's atmospheric extinction still exist. Typical atmospheric extinction curves are smooth functions of wavelength in the LAMOST wavelength coverage \citep{bongard, cullen}, and this is usually true for the variations of atmospheric extinction, which  can be well represented by low order polynomials.  Therefore, the mean
atmospheric extinction curve included in the ASPSRCs does not affect the spectral lines of the calibrated spectra. Its variations are included in the overall variations of SRCs.

Secondly,  fiber positioning may introduce variations in the fiber spectral response during  the tracing  of the targets.  This means the SRCs of the individual fibers probably vary from observations of one plate to another \citep{chen}. The variations of the fiber flat fields will  have an impact on sky subtraction and flux calibration, introducing the uncertainties to the SRCs. To make matters worse uncertainties introduced by such variations  do not depend on the spectral SNRs. That is to say, even spectra of very high SNRs may have incorrectly shaped SEDs. Attempts to characterize and correct for such variations of  the fiber flat fields is under way. If the condition allows, it is better to obtain the ASPSRCs once a year or once a quarter to overcome these instrumental changes.

Thirdly, the spectral SNRs of standard stars  have an impact on the SRCs derived from them.  To test how the SEDs of LAMOST Galactic targets are affected by limited SNRs, the spectral and photometric (g$-$r) colors have been compared as a function of the spectral SNRs.  The results shows that at SNRs  exceeding 20,  the spectral colors and photometric colors agree well, with a mean difference of 0.01-0.02 mag and no systematic trend, while at SNRs lower than about 10, the discrepancies increase rapidly, along with some systematic differences \citep{xiang}. To minimize the uncertainties introduced by spectral SNRs, we selected the standard stars with SNRs larger than 20 to obtain the ASPSRCs.

In addition, errors due to the stellar atmospheric parameters of standard stars also cause variations in the SRCs.  For flux standard stars of  5,750 K$\leq T_{\rm eff} \leq$6,750 K, an error of 150 K in $T_{\rm eff}$ can lead to a maximum uncertainty of 12\%  in the shape of the stellar SED and thus it will change the shape of the SRC derived from it. Uncertainties caused by errors in \textbf{log g} are negligible  (i.e. for an estimated  uncertainty of 0.25 dex in log g, about 1\%  for the whole wavelength range  is affected).  Metallicity mainly affects the blue-arm spectra at wavelengths less than 4,500 \AA.  An error of 0.2  dex in [Fe/H] can change the SED shape between 3,800 \AA \ and 4,500 \AA \ by approximately 3\%, while the effects at wavelengths greater than 4, 500 \AA \ are only marginally \citep{xiang}.  This is the reason we remove the candidates with standards which above uncertainties in $T_{\rm eff}$ larger than 150 K.

The advantage of the ASPSRCs comes from using an average SRC of instrument response curves and the average of atmospheric extinction curves, although there are many uncertainties introduced by the influencing factors discussed, which will eliminate the effects. Our experiments prove that all the influencing factors on accuracy of flux calibration is  less than 10\ during the DR2 period. The average SRCs are presented in Table \ref{tab3} to Table \ref{tab6}.  One can use them to calibrate spectra of the LAMOST DR2 catalogue.  For the spectra observed subsequent to DR2, new ASPSRCs will need to be produced to counter variations from the instrument.

\section*{Acknowledgements}

The authors would like to thank M.--S. Xiang, H.--B. Yuan for helpful discussion. This work is supported by the by the National Key Basic Research Program of China (Grant No.2014CB845700), and National Science Foundation of China (Grant Nos 11390371, 11233004).
Guoshoujing Telescope (the Large Sky Area Multi-Object Fiber Spectroscopic Telescope, LAMOST) is a National Major Scientific Project built by the Chinese Academy of Sciences. Funding for the project has been provided by the National Development and Reform Commission. LAMOST is operated and managed by the National Astronomical Observatories, Chinese Academy of Sciences.

\clearpage

\clearpage

\clearpage

\clearpage

\clearpage
\clearpage

\clearpage
\clearpage

\clearpage
\clearpage

\clearpage
\clearpage

\clearpage

\clearpage

\clearpage

\clearpage

\clearpage

\clearpage

\clearpage

\begin{figure}
\centering
\includegraphics[width=18.0cm,height=18.0cm]{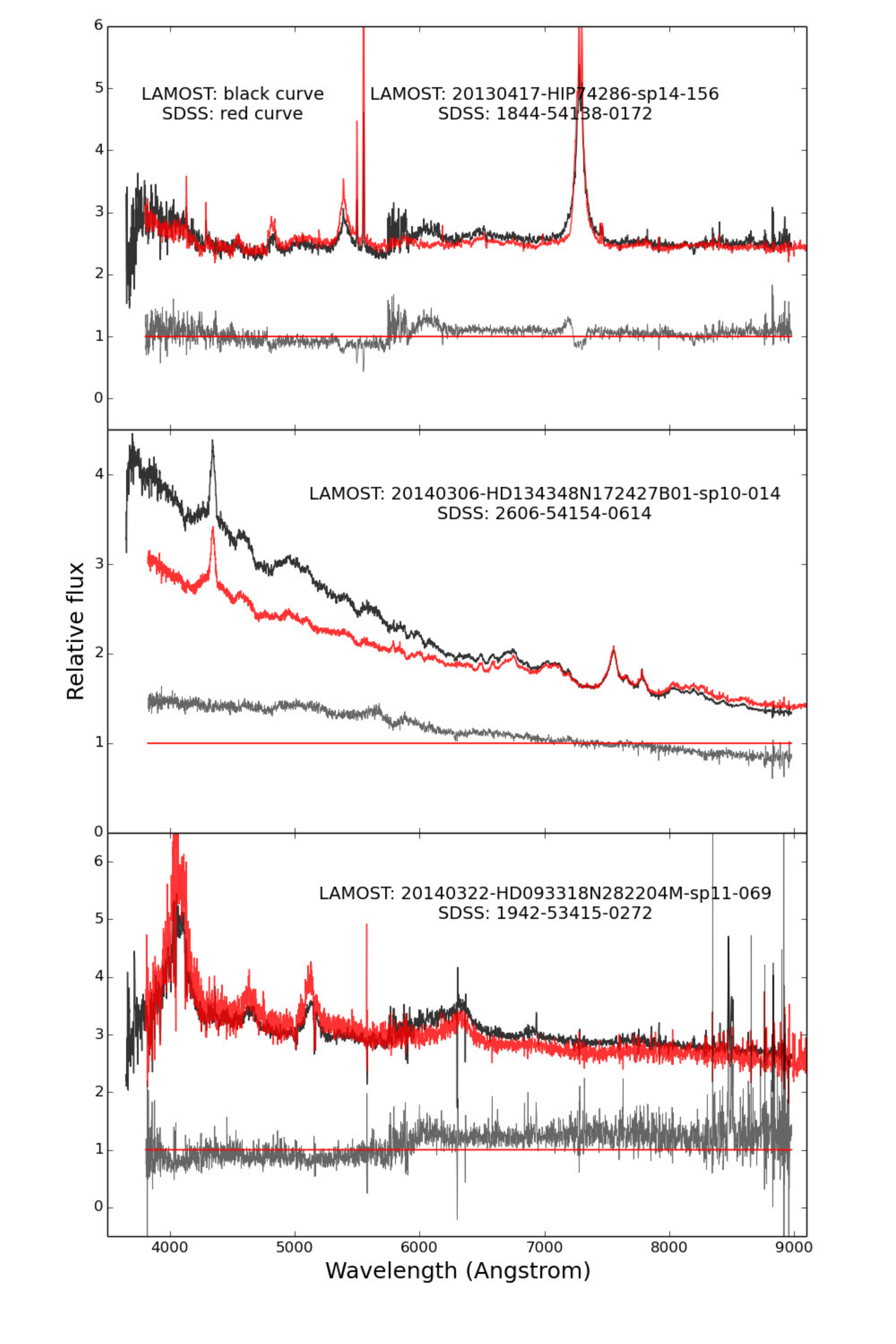}
\caption{Comparison of the rescued spectra of QSOs (black) with SDSS DR12 spectra (red).}
\label{fig16}
\end{figure}
\clearpage

\begin{figure}
\centering
\includegraphics[width=18.0cm,height=18.0cm]{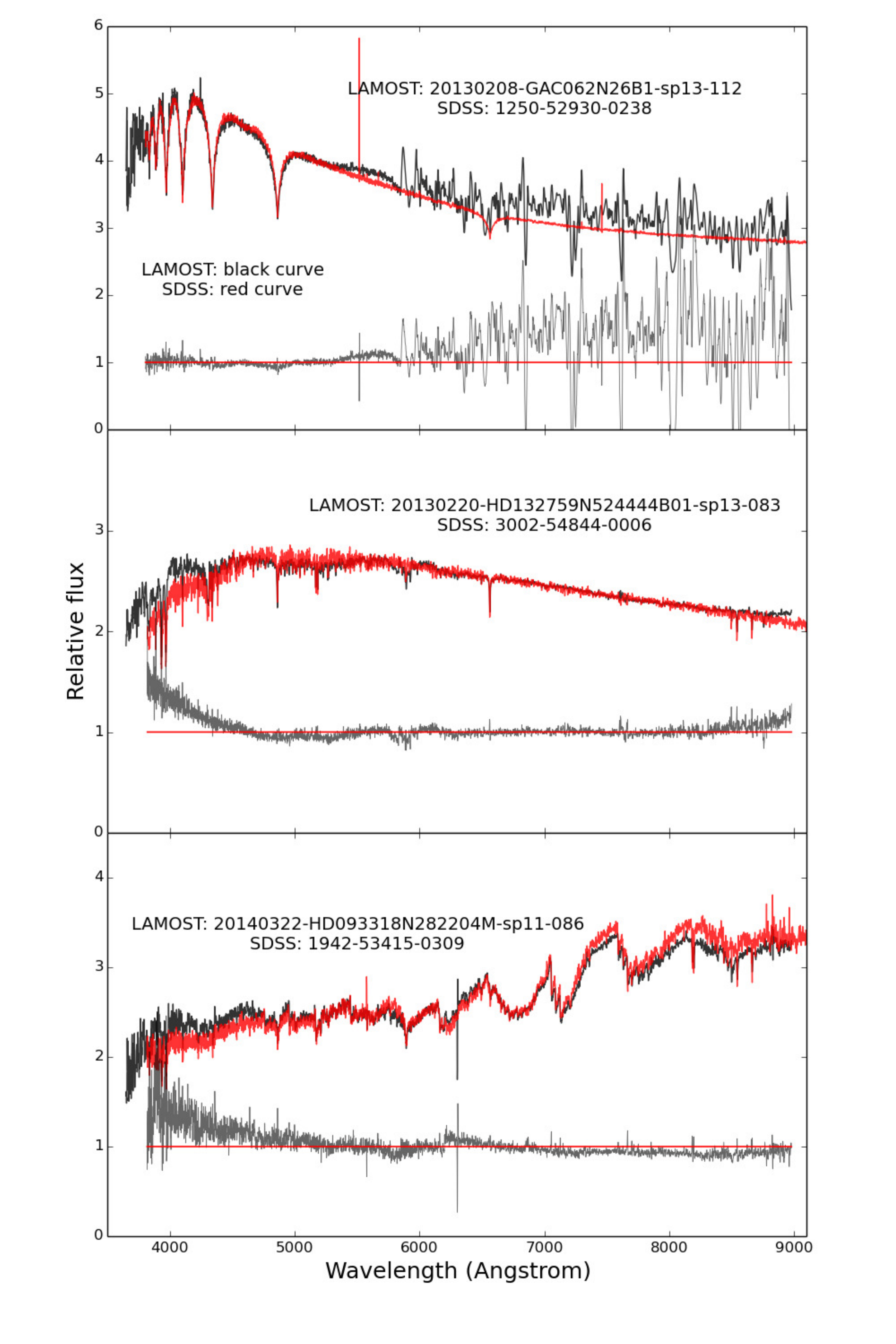}
\caption{Comparison of the rescued spectra of stars (black) with SDSS DR12 spectra (red). }
\label{fig17}
\end{figure}
\clearpage

\begin{figure}
\centering
\includegraphics[width=15.0cm]{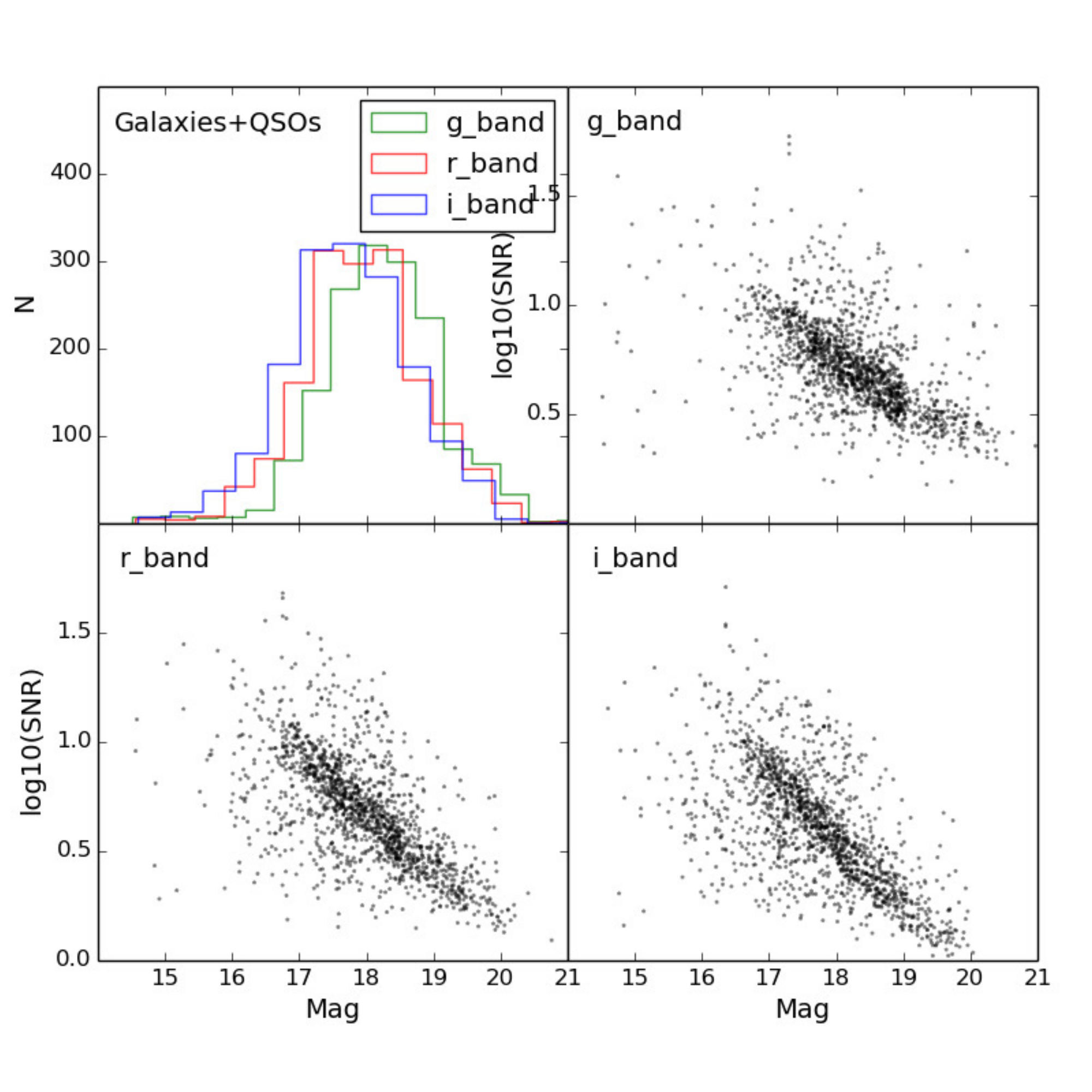}
\caption{Histogram of the g,r,i magnitudes of galaxies and QSOs, which are rescued by  the ASPSRCs from  the abandoned spectrographs of  the 2D pipeline.  The diagrams of SNRs and magnitudes are provided for g,r,i--bands. }
\label{fig18}
\end{figure}
\clearpage

\begin{figure}
\centering
\includegraphics[width=15.0cm]{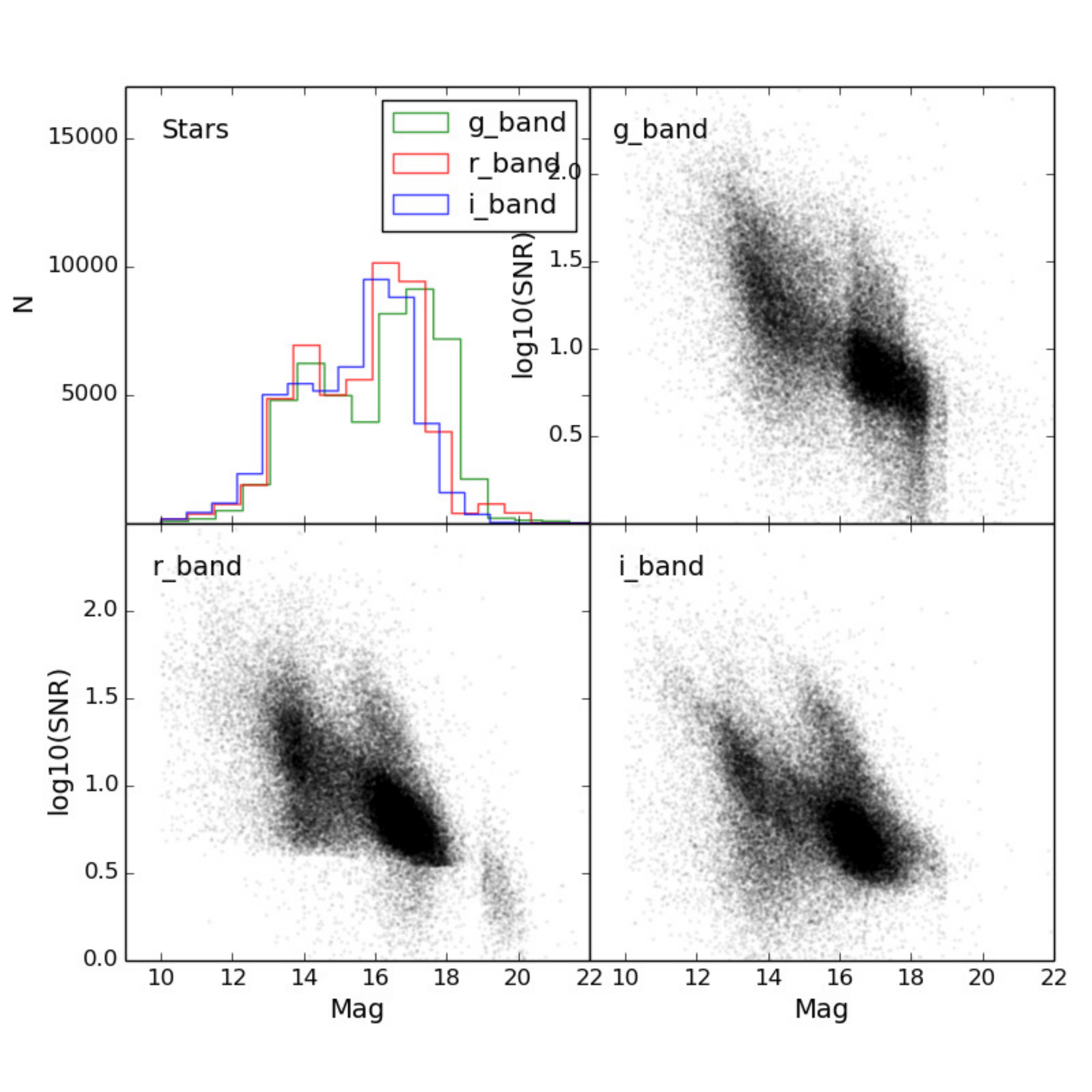}
\caption{Histogram of the g,r,i magnitudes of  stars, which  are rescued by  the ASPSRCs from the abandoned spectrographs of the 2D pipeline.  The diagrams of SNRs and magnitudes are provided for g,r,i--bands. }
\label{fig19}
\end{figure}
\clearpage

\begin{figure}
\centering
\includegraphics[width=15.0cm]{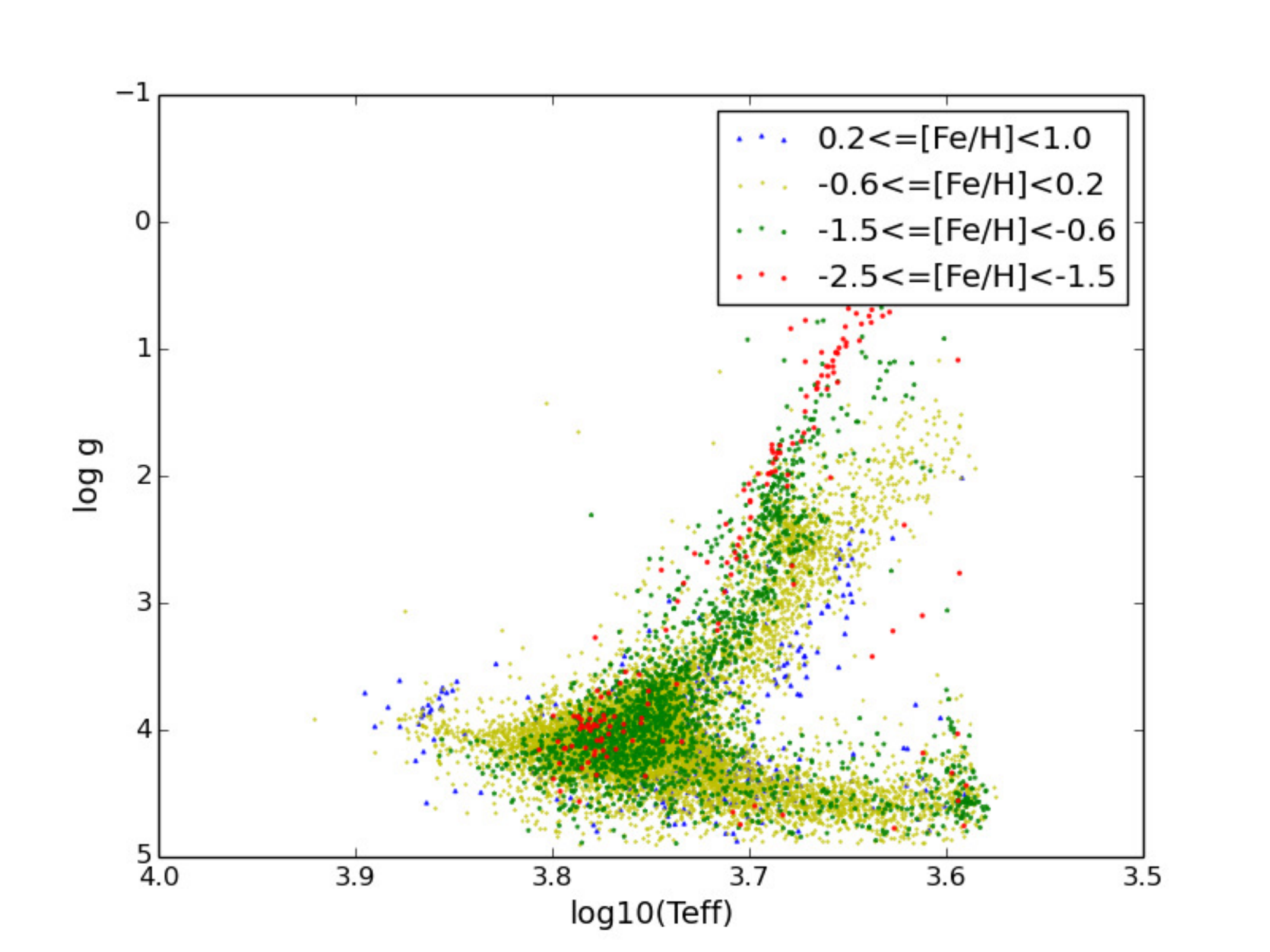}
\caption{The coverage map of stellar parameters of F,G,K--type  stars.}
\label{fig20}
\end{figure}
\clearpage

\begin{figure}
\centering
\includegraphics[width=15.0cm]{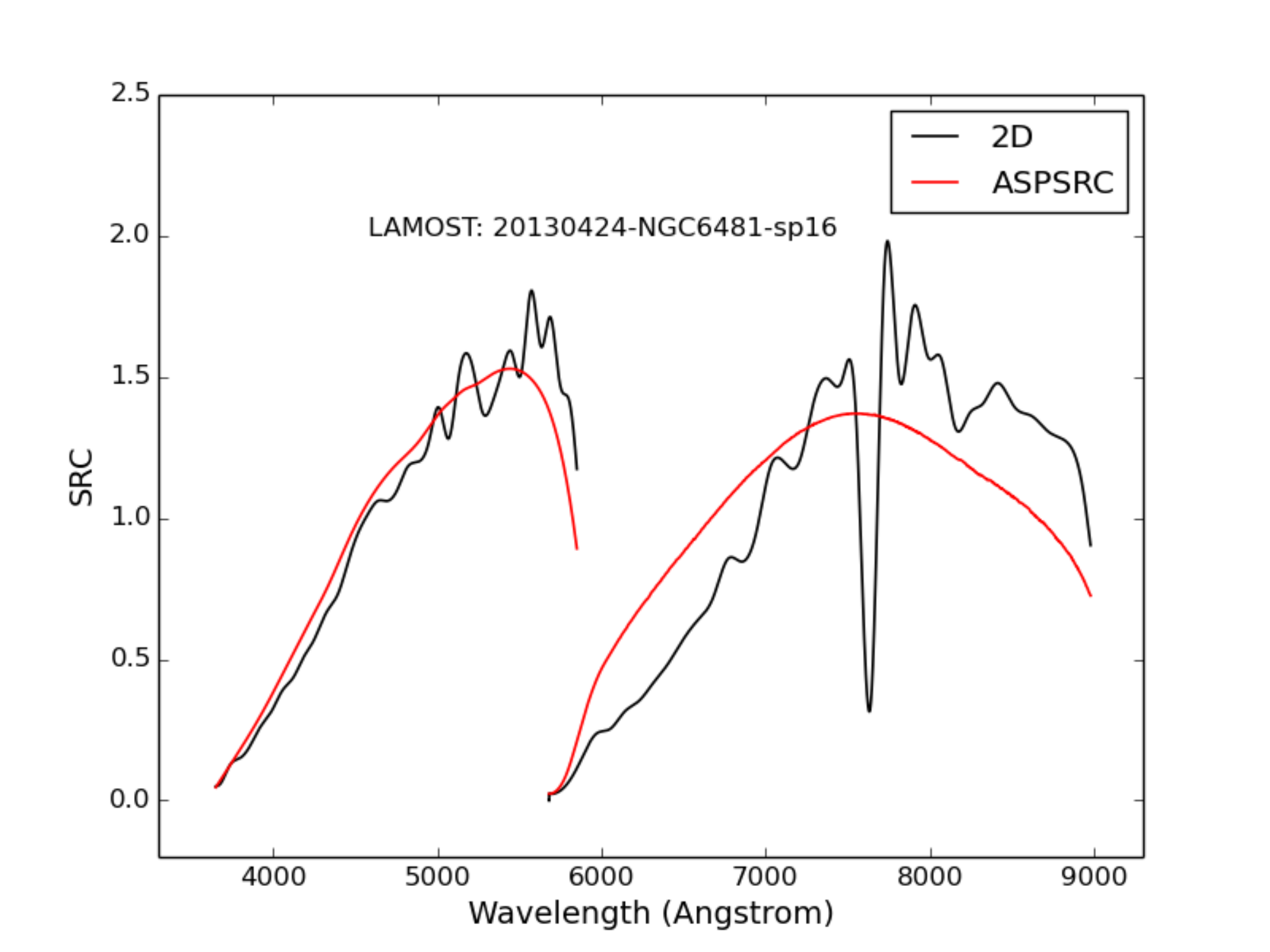}
\caption{Comparison of the over fitted SRCs from the 2D pipeline (black) with the ASPSRCs (red).}
\label{fig21}
\end{figure}
\clearpage

\begin{figure}
\centering
\includegraphics[width=15.0cm]{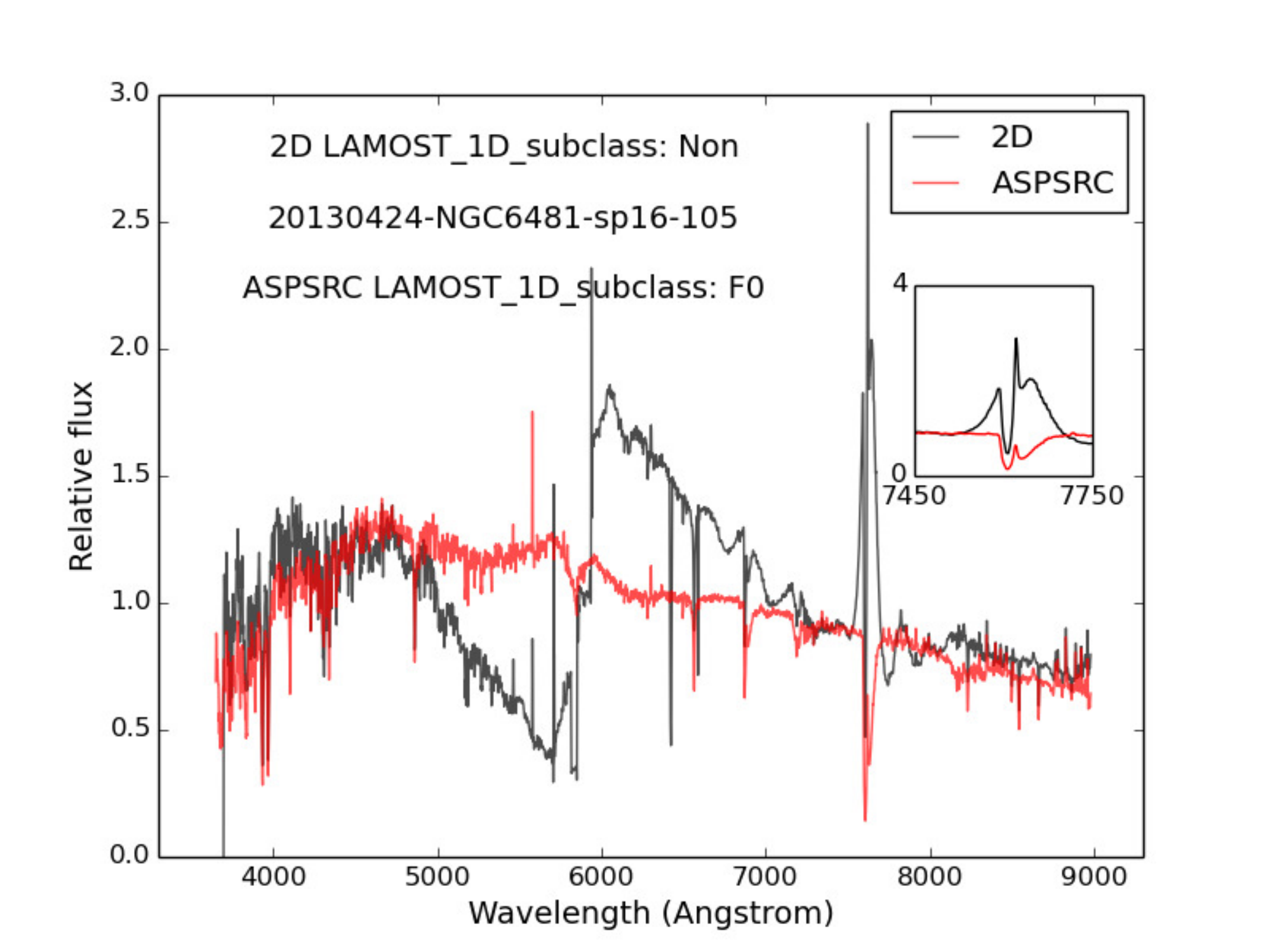}
\caption{Comparison of the artificial spectra calibrated by adopting the over fitted SRCs in Fig \ref{fig18} (black) with the spectra recalibrated by the ASPSRCs (red). The spectra is classified as an F0 instead of "Non"  after recalibration.}
\label{fig22}
\end{figure}
\clearpage
\begin{table}
\begin{center}
\caption[]{The absolute and relative uncertainties of the g r i regions.}\label{tab1}
%%Please Capitalize the First Letter of Each Notional Word in table's caption
 \begin{tabular}{lllllll}
 %\hline\noalign{\smallskip}
 No-spectrograph & g-absolute & g-relative & r-absolute & r-relative &i-absolute &i-relative \\
 \hline\noalign{\smallskip} \hline\noalign{\smallskip}
sp01 &0.054 &6.35\% &0.054 &6.75\% &0.033 & 2.83\% \\ % new variable
sp02 &0.046 &5.12\% &0.045 &5.96\% &0.028 & 2.41\% \\ % new variable
sp03 &0.047 &4.71\% &0.033 &6.34\% &0.031 &2.74\%  \\
sp04 &0.048 &5.49\% &0.039 &6.74\% &0.031 &2.72\% \\
sp05 &0.053 &6.95\% &0.065 &11.13\% &0.040 &3.51\% \\
sp06 &0.051 &5.07\% &0.049 &7.83\% &0.035 &3.20\% \\
sp07 &0.048 &5.22\% &0.046 &6.21\% &0.031 &2.71\% \\
sp08 &0.052 &6.30\% &0.052 &6.91\% &0.038 &3.38\% \\
sp09 &0.053 &6.57\% &0.051 &6.46\% &0.031 &2.74\% \\
sp10 &0.046	&5.25\%	&0.046 &7.32\% &0.039 &3.60\% \\	
sp11 &0.045	&5.57\%	&0.036 &6.70\% &0.033 &2.93\% \\	
sp12 &0.054	&6.71\%	&0.047 &7.88\% &0.047 &4.11\%	\\
sp13 &0.053	&6.44\%	&0.041 &6.94\% &0.034 &2.99\%  \\	
sp14 &0.048	&5.97\%	&0.039 &6.74\% &0.033 &2.83\%  \\	
sp15 &0.041	&5.12\%	&0.035 &5.82\% &0.029 &2.60\%   \\	
sp16 &0.041	&5.02\%	&0.034 &5.66\% &0.032 &2.82\%  \\	
\noalign{\smallskip}\hline
\end{tabular}
\end{center}
\end{table}

\clearpage
\begin{table}
\begin{center}
\caption[]{The final released spectral classifications of the abandoned spectrographs.}\label{tab2}
%%Please Capitalize the First Letter of Each Notional Word in table's caption
 \begin{tabular}{ll}
 %\hline\noalign{\smallskip}
 Type &Number \\
 \hline\noalign{\smallskip} \hline\noalign{\smallskip}
 total       52181  \\
 Galaxy      &1163    \\
 QSO         &201  \\
 BA           &1477 \\
 F           &8454  \\
 G           &13875 \\
 K           &14440  \\
 M           &12571  \\
\noalign{\smallskip}\hline
\end{tabular}
\end{center}
\end{table}
\clearpage

%\begin{landscape}
\tiny
\begin{longtable}[]{lllllllll}
\caption{The blue--arm ASPSRCs of LAMOST spectrographs from No.1 to No.8 at 100 \AA steps.  All the ASPSRCs are scaled to a mean value of unity. } \label{tab3} \\

%\multicolumn{1}{l}{} & \multicolumn{1}{l}{ No.1} & \multicolumn{1}{l}{No.2 } & \multicolumn{1}{l}{No.3} & \multicolumn{1}{l}{No.4 } &
%\multicolumn{1}{l}{No.5} & \multicolumn{1}{l}{ No.6 } & \multicolumn{1}{l}{No.7 } & \multicolumn{1}{l}{ No.8 }\\
%\multicolumn{1}{l}{Wavelength} & \multicolumn{1}{l}{SRC(Err)} & \multicolumn{1}{l}{ SRC(Err)} & \multicolumn{1}{l}{SRC(Err)} & \multicolumn{1}{l}{ SRC(Err) } &
%\multicolumn{1}{l}{ SRC(Err)} & \multicolumn{1}{l}{ SRC(Err)} & \multicolumn{1}{l}{ SRC(Err)} & \multicolumn{1}{l}{SRC(Err)}\\
           &No.1     &No.2      &No.3     &No.4      &No.5     &No.6      &No.7      &No.8 \\
Wavelength &SRC(Err) &SRC(Err)  &SRC(Err) &SRC(Err)  &SRC(Err) &SRC(Err)  &SRC(Err)  &SRC(Err) \\
\hline \hline
\endfirsthead
\multicolumn{9}{r}%
{{\bfseries \tablename\ \thetable{} -- continued from previous page}} \\
           &No.1     &No.2      &No.3     &No.4      &No.5     &No.6      &No.7      &No.8 \\
Wavelength &SRC(Err) &SRC(Err)  &SRC(Err) &SRC(Err)  &SRC(Err) &SRC(Err)  &SRC(Err)  &SRC(Err) \\
\hline \hline
\endhead
\hline \multicolumn{9}{r}{{Continued on next page}} \\
\endfoot
\endlastfoot

3650 &0.142(0.036) &0.132(0.027) &0.106(0.016) &0.070(0.026) &0.054(0.054) &0.150(0.027) &0.098(0.016) &0.048(0.053) \\
3750 &0.245(0.051) &0.244(0.039) &0.261(0.040) &0.187(0.030) &0.122(0.046) &0.268(0.044) &0.202(0.034) &0.156(0.029) \\
3850 &0.351(0.063) &0.356(0.048) &0.409(0.051) &0.307(0.041) &0.207(0.035) &0.397(0.056) &0.325(0.049) &0.267(0.039) \\
3950 &0.460(0.070) &0.467(0.054) &0.550(0.058) &0.429(0.050) &0.305(0.043) &0.531(0.065) &0.455(0.058) &0.379(0.048) \\
4050 &0.570(0.076) &0.578(0.060) &0.682(0.063) &0.551(0.059) &0.415(0.069) &0.670(0.072) &0.586(0.068) &0.491(0.059) \\
4150 &0.680(0.077) &0.687(0.061) &0.805(0.065) &0.671(0.062) &0.534(0.053) &0.808(0.073) &0.716(0.072) &0.602(0.063) \\
4250 &0.789(0.075) &0.795(0.059) &0.917(0.066) &0.789(0.063) &0.659(0.071) &0.943(0.072) &0.844(0.072) &0.710(0.064) \\
4350 &0.896(0.069) &0.900(0.054) &1.022(0.061) &0.903(0.060) &0.788(0.084) &1.073(0.067) &0.973(0.064) &0.820(0.064) \\
4450 &1.005(0.064) &1.007(0.050) &1.124(0.057) &1.010(0.059) &0.919(0.054) &1.197(0.060) &1.110(0.057) &0.942(0.063) \\
4550 &1.120(0.056) &1.130(0.043) &1.224(0.049) &1.110(0.054) &1.050(0.051) &1.302(0.048) &1.251(0.048) &1.072(0.058) \\
4650 &1.190(0.046) &1.207(0.037) &1.294(0.041) &1.201(0.047) &1.174(0.045) &1.374(0.038) &1.359(0.038) &1.157(0.053) \\
4750 &1.270(0.037) &1.288(0.033) &1.339(0.031) &1.281(0.036) &1.281(0.037) &1.446(0.031) &1.476(0.028) &1.277(0.044) \\
4850 &1.353(0.026) &1.400(0.033) &1.363(0.022) &1.344(0.022) &1.363(0.032) &1.493(0.028) &1.565(0.016) &1.396(0.032) \\
4950 &1.423(0.016) &1.511(0.034) &1.394(0.015) &1.372(0.013) &1.434(0.036) &1.514(0.027) &1.634(0.014) &1.492(0.019) \\
5050 &1.484(0.022) &1.604(0.031) &1.432(0.020) &1.417(0.018) &1.507(0.030) &1.533(0.029) &1.702(0.026) &1.572(0.014) \\
5150 &1.538(0.046) &1.667(0.032) &1.429(0.030) &1.456(0.028) &1.577(0.038) &1.554(0.039) &1.782(0.040) &1.683(0.031) \\
5250 &1.584(0.065) &1.705(0.042) &1.383(0.039) &1.477(0.038) &1.608(0.050) &1.543(0.051) &1.788(0.056) &1.775(0.050) \\
5350 &1.618(0.082) &1.721(0.064) &1.336(0.050) &1.472(0.047) &1.613(0.062) &1.515(0.064) &1.773(0.073) &1.814(0.071) \\
5450 &1.626(0.091) &1.685(0.085) &1.280(0.059) &1.442(0.057) &1.605(0.077) &1.455(0.073) &1.681(0.087) &1.797(0.088) \\
5550 &1.564(0.090) &1.536(0.096) &1.185(0.067) &1.394(0.067) &1.571(0.099) &1.303(0.079) &1.331(0.088) &1.683(0.093) \\
5650 &1.340(0.084) &1.205(0.092) &1.068(0.070) &1.329(0.074) &1.498(0.111) &0.979(0.085) &0.769(0.064) &1.409(0.080) \\
5750 &0.845(0.082) &0.604(0.076) &0.928(0.068) &1.229(0.077) &1.351(0.097) &0.430(0.074) &0.257(0.028) &0.853(0.054) \\
5850 &0.114(0.017) &0.095(0.017) &0.749(0.063) &1.060(0.075) &0.977(0.272) &0.134(0.034) &0.007(0.000) &0.127(0.009) \\

\noalign{\smallskip}\hline
\end{longtable}
%\end{landscape}
%\endgroup

\clearpage

%\begin{landscape}
\tiny
\begin{longtable}[]{lllllllll}
\caption{The red--arm ASPSRCs of LAMOST spectrographs from No.1 to No.8 at 100 \AA steps.} \label{tab4} \\

%\multicolumn{1}{l}{} & \multicolumn{1}{l}{ No.1} & \multicolumn{1}{l}{No.2 } & \multicolumn{1}{l}{No.3} & \multicolumn{1}{l}{No.4 } &
%\multicolumn{1}{l}{No.5} & \multicolumn{1}{l}{ No.6 } & \multicolumn{1}{l}{No.7 } & \multicolumn{1}{l}{ No.8 }\\
%\multicolumn{1}{l}{Wavelength} & \multicolumn{1}{l}{SRC(Err)} & \multicolumn{1}{l}{ SRC(Err)} & \multicolumn{1}{l}{SRC(Err)} & \multicolumn{1}{l}{ SRC(Err) } &
%\multicolumn{1}{l}{ SRC(Err)} & \multicolumn{1}{l}{ SRC(Err)} & \multicolumn{1}{l}{ SRC(Err)} & \multicolumn{1}{l}{SRC(Err)}\\
           &No.1     &No.2      &No.3     &No.4      &No.5     &No.6      &No.7      &No.8 \\
Wavelength &SRC(Err) &SRC(Err)  &SRC(Err) &SRC(Err)  &SRC(Err) &SRC(Err)  &SRC(Err)  &SRC(Err) \\
\hline \hline
\endfirsthead
\multicolumn{9}{r}%
{{\bfseries \tablename\ \thetable{} -- continued from previous page}} \\
           &No.1     &No.2      &No.3     &No.4      &No.5     &No.6      &No.7      &No.8 \\
Wavelength &SRC(Err) &SRC(Err)  &SRC(Err) &SRC(Err)  &SRC(Err) &SRC(Err)  &SRC(Err)  &SRC(Err) \\
\hline \hline
\endhead
\hline \multicolumn{9}{r}{{Continued on next page}} \\
\endfoot
\endlastfoot

5680 &0.070(0.047) &0.116(0.030) &-0.005(0.003) &0.006(0.009) &-0.018(0.008) &0.185(0.057) &0.329(0.065) &0.081(0.032) \\
5780 &0.333(0.047) &0.330(0.045) &0.031(0.007) &0.044(0.007) &0.074(0.070) &0.269(0.049) &0.433(0.049) &0.273(0.035) \\
5880 &0.509(0.062) &0.478(0.049) &0.140(0.014) &0.187(0.020) &0.273(0.088) &0.356(0.053) &0.520(0.053) &0.423(0.056) \\
5980 &0.620(0.065) &0.569(0.046) &0.283(0.028) &0.371(0.044) &0.458(0.091) &0.436(0.054) &0.601(0.053) &0.537(0.065) \\
6080 &0.696(0.064) &0.639(0.046) &0.427(0.039) &0.545(0.061) &0.585(0.088) &0.513(0.057) &0.674(0.054) &0.626(0.069) \\
6180 &0.760(0.062) &0.711(0.047) &0.543(0.044) &0.675(0.066) &0.672(0.079) &0.587(0.059) &0.748(0.054) &0.703(0.069) \\
6280 &0.820(0.056) &0.779(0.046) &0.634(0.046) &0.763(0.065) &0.745(0.077) &0.659(0.059) &0.817(0.053) &0.770(0.068) \\
6380 &0.882(0.053) &0.855(0.046) &0.715(0.049) &0.834(0.062) &0.815(0.078) &0.735(0.061) &0.892(0.054) &0.842(0.067) \\
6480 &0.938(0.048) &0.929(0.044) &0.791(0.050) &0.895(0.058) &0.881(0.070) &0.810(0.059) &0.960(0.052) &0.905(0.063) \\
6580 &0.995(0.044) &1.001(0.042) &0.867(0.048) &0.956(0.053) &0.949(0.064) &0.884(0.057) &1.024(0.049) &0.970(0.059) \\
6680 &1.052(0.041) &1.071(0.039) &0.943(0.046) &1.019(0.047) &1.020(0.062) &0.956(0.054) &1.086(0.047) &1.035(0.055) \\
6780 &1.104(0.038) &1.137(0.035) &1.019(0.044) &1.080(0.041) &1.086(0.051) &1.026(0.049) &1.141(0.043) &1.100(0.050) \\
6880 &1.154(0.034) &1.198(0.033) &1.091(0.041) &1.138(0.034) &1.151(0.042) &1.091(0.044) &1.192(0.037) &1.161(0.046) \\
6980 &1.196(0.033) &1.250(0.028) &1.157(0.038) &1.190(0.032) &1.207(0.034) &1.152(0.036) &1.235(0.032) &1.212(0.038) \\
7080 &1.234(0.032) &1.297(0.024) &1.219(0.034) &1.239(0.026) &1.258(0.031) &1.207(0.029) &1.273(0.026) &1.261(0.033) \\
7180 &1.265(0.029) &1.334(0.021) &1.273(0.030) &1.279(0.022) &1.300(0.031) &1.254(0.022) &1.302(0.021) &1.298(0.028) \\
7280 &1.288(0.029) &1.362(0.019) &1.320(0.027) &1.312(0.021) &1.332(0.030) &1.291(0.019) &1.323(0.018) &1.328(0.025) \\
7380 &1.304(0.025) &1.379(0.014) &1.357(0.018) &1.336(0.013) &1.353(0.024) &1.319(0.012) &1.333(0.010) &1.349(0.021) \\
7480 &1.312(0.025) &1.384(0.013) &1.384(0.014) &1.350(0.015) &1.366(0.022) &1.339(0.012) &1.336(0.010) &1.361(0.019) \\
7580 &1.313(0.023) &1.378(0.015) &1.401(0.011) &1.355(0.020) &1.366(0.024) &1.349(0.015) &1.329(0.012) &1.361(0.019) \\
7680 &1.305(0.024) &1.363(0.017) &1.407(0.014) &1.350(0.022) &1.356(0.028) &1.351(0.020) &1.313(0.018) &1.348(0.022) \\
7780 &1.291(0.023) &1.339(0.021) &1.407(0.018) &1.338(0.026) &1.337(0.028) &1.345(0.024) &1.287(0.021) &1.327(0.023) \\
7880 &1.271(0.024) &1.308(0.021) &1.399(0.018) &1.318(0.026) &1.313(0.030) &1.331(0.029) &1.255(0.025) &1.298(0.026) \\
7980 &1.246(0.028) &1.269(0.025) &1.385(0.022) &1.291(0.027) &1.286(0.032) &1.313(0.034) &1.218(0.029) &1.261(0.031) \\
8080 &1.218(0.033) &1.230(0.030) &1.366(0.025) &1.264(0.029) &1.252(0.041) &1.292(0.041) &1.178(0.034) &1.216(0.038) \\
8180 &1.180(0.035) &1.186(0.036) &1.342(0.031) &1.232(0.030) &1.220(0.046) &1.267(0.045) &1.130(0.040) &1.171(0.050) \\
8280 &1.139(0.041) &1.139(0.043) &1.313(0.037) &1.198(0.031) &1.187(0.052) &1.235(0.053) &1.082(0.044) &1.120(0.061) \\
8380 &1.101(0.041) &1.092(0.042) &1.286(0.043) &1.170(0.038) &1.158(0.047) &1.203(0.052) &1.032(0.046) &1.081(0.056) \\
8480 &1.062(0.045) &1.042(0.048) &1.256(0.047) &1.142(0.044) &1.128(0.046) &1.163(0.053) &0.980(0.049) &1.034(0.061) \\
8580 &1.014(0.045) &0.986(0.050) &1.222(0.051) &1.117(0.045) &1.091(0.048) &1.116(0.055) &0.925(0.051) &0.977(0.065) \\
8680 &0.962(0.046) &0.925(0.051) &1.182(0.054) &1.091(0.047) &1.048(0.053) &1.060(0.053) &0.868(0.052) &0.921(0.068) \\
8780 &0.907(0.046) &0.853(0.049) &1.125(0.055) &1.047(0.050) &0.990(0.053) &0.991(0.051) &0.806(0.051) &0.855(0.067) \\
8880 &0.841(0.050) &0.773(0.049) &1.037(0.059) &0.972(0.054) &0.916(0.056) &0.911(0.052) &0.740(0.054) &0.778(0.064) \\
8980 &0.757(0.044) &0.677(0.044) &0.898(0.053) &0.840(0.043) &0.794(0.060) &0.820(0.052) &0.670(0.051) &0.673(0.069) \\

\noalign{\smallskip}\hline
\end{longtable}
%\end{landscape}

\clearpage

%\begin{landscape}
\tiny
\begin{longtable}[]{lllllllll}
\caption{The blue--arm ASPSRCs of LAMOST spectrographs from No.9 to No.16 at 100 \AA steps.} \label{tab5} \\

%\multicolumn{1}{l}{} & \multicolumn{1}{l}{ No.1} & \multicolumn{1}{l}{No.2 } & \multicolumn{1}{l}{No.3} & \multicolumn{1}{l}{No.4 } &
%\multicolumn{1}{l}{No.5} & \multicolumn{1}{l}{ No.6 } & \multicolumn{1}{l}{No.7 } & \multicolumn{1}{l}{ No.8 }\\
%\multicolumn{1}{l}{Wavelength} & \multicolumn{1}{l}{SRC(Err)} & \multicolumn{1}{l}{ SRC(Err)} & \multicolumn{1}{l}{SRC(Err)} & \multicolumn{1}{l}{ SRC(Err) } &
%\multicolumn{1}{l}{ SRC(Err)} & \multicolumn{1}{l}{ SRC(Err)} & \multicolumn{1}{l}{ SRC(Err)} & \multicolumn{1}{l}{SRC(Err)}\\
           &No.9     &No.10      &No.11     &No.12      &No.13     &No.14      &No.15      &No.16 \\
Wavelength &SRC(Err) &SRC(Err)  &SRC(Err) &SRC(Err)  &SRC(Err) &SRC(Err)  &SRC(Err)  &SRC(Err) \\
\hline \hline
\endfirsthead
\multicolumn{9}{r}%
{{\bfseries \tablename\ \thetable{} -- continued from previous page}} \\
           &No.9     &No.10      &No.11     &No.12      &No.13     &No.14      &No.15      &No.16 \\
Wavelength &SRC(Err) &SRC(Err)  &SRC(Err) &SRC(Err)  &SRC(Err) &SRC(Err)  &SRC(Err)  &SRC(Err) \\
\hline \hline
\endhead
\hline \multicolumn{9}{r}{{Continued on next page}} \\
\endfoot
\endlastfoot

3650 &0.080(0.012) &0.096(0.019) &0.082(0.014) &0.027(0.006) &0.058(0.027) &0.069(0.023) &0.013(0.037) &0.049(0.018) \\
3750 &0.160(0.027) &0.201(0.036) &0.179(0.025) &0.122(0.019) &0.106(0.017) &0.132(0.028) &0.154(0.031) &0.144(0.020) \\
3850 &0.254(0.038) &0.309(0.046) &0.275(0.033) &0.228(0.032) &0.171(0.023) &0.199(0.033) &0.272(0.040) &0.241(0.029) \\
3950 &0.359(0.046) &0.428(0.054) &0.375(0.039) &0.338(0.042) &0.249(0.031) &0.277(0.041) &0.390(0.045) &0.347(0.036) \\
4050 &0.470(0.056) &0.557(0.061) &0.480(0.046) &0.449(0.051) &0.338(0.039) &0.372(0.048) &0.509(0.051) &0.463(0.044) \\
4150 &0.581(0.063) &0.686(0.064) &0.587(0.051) &0.559(0.056) &0.435(0.042) &0.479(0.052) &0.622(0.053) &0.583(0.049) \\
4250 &0.690(0.068) &0.802(0.066) &0.694(0.054) &0.666(0.063) &0.536(0.045) &0.589(0.055) &0.716(0.053) &0.699(0.051) \\
4350 &0.805(0.067) &0.923(0.062) &0.806(0.056) &0.779(0.064) &0.649(0.048) &0.709(0.055) &0.801(0.049) &0.822(0.051) \\
4450 &0.936(0.067) &1.066(0.058) &0.930(0.057) &0.908(0.065) &0.782(0.054) &0.844(0.055) &0.908(0.046) &0.961(0.051) \\
4550 &1.061(0.062) &1.185(0.050) &1.045(0.054) &1.029(0.065) &0.912(0.059) &0.973(0.053) &1.013(0.041) &1.084(0.046) \\
4650 &1.161(0.051) &1.272(0.040) &1.136(0.048) &1.138(0.060) &1.033(0.061) &1.085(0.049) &1.104(0.036) &1.181(0.040) \\
4750 &1.247(0.040) &1.344(0.029) &1.209(0.039) &1.235(0.052) &1.148(0.061) &1.186(0.046) &1.184(0.031) &1.255(0.032) \\
4850 &1.346(0.029) &1.417(0.018) &1.284(0.030) &1.319(0.040) &1.265(0.056) &1.293(0.042) &1.265(0.027) &1.314(0.026) \\
4950 &1.444(0.026) &1.497(0.015) &1.369(0.022) &1.383(0.027) &1.400(0.049) &1.399(0.040) &1.354(0.023) &1.395(0.020) \\
5050 &1.522(0.033) &1.558(0.016) &1.439(0.013) &1.443(0.012) &1.517(0.038) &1.492(0.036) &1.443(0.022) &1.473(0.017) \\
5150 &1.574(0.048) &1.597(0.027) &1.472(0.017) &1.508(0.016) &1.586(0.025) &1.561(0.033) &1.512(0.024) &1.528(0.018) \\
5250 &1.615(0.054) &1.592(0.039) &1.475(0.026) &1.536(0.034) &1.636(0.025) &1.590(0.034) &1.539(0.032) &1.558(0.025) \\
5350 &1.660(0.062) &1.606(0.055) &1.478(0.040) &1.536(0.054) &1.697(0.037) &1.605(0.039) &1.570(0.040) &1.597(0.037) \\
5450 &1.682(0.073) &1.595(0.069) &1.474(0.054) &1.528(0.070) &1.736(0.056) &1.614(0.049) &1.576(0.049) &1.612(0.050) \\
5550 &1.619(0.088) &1.497(0.081) &1.441(0.067) &1.508(0.081) &1.705(0.076) &1.591(0.061) &1.526(0.057) &1.586(0.064) \\
5650 &1.431(0.094) &1.289(0.088) &1.385(0.079) &1.472(0.088) &1.634(0.092) &1.545(0.073) &1.442(0.062) &1.501(0.075) \\
5750 &1.069(0.087) &0.873(0.075) &1.295(0.088) &1.403(0.095) &1.514(0.106) &1.456(0.081) &1.305(0.066) &1.308(0.078) \\
5850 &0.524(0.060) &0.168(0.019) &1.147(0.089) &1.283(0.097) &1.302(0.111) &1.290(0.086) &1.078(0.067) &0.941(0.068) \\
\noalign{\smallskip}\hline
\end{longtable}
%\end{landscape}

\clearpage

\tiny
\begin{longtable}[]{lllllllll}
\caption{The red--arm  ASPSRCs of LAMOST spectrographs from No.9 to No.16 at 100 \AA steps.} \label{tab6} \\

%\multicolumn{1}{l}{} & \multicolumn{1}{l}{ No.1} & \multicolumn{1}{l}{No.2 } & \multicolumn{1}{l}{No.3} & \multicolumn{1}{l}{No.4 } &
%\multicolumn{1}{l}{No.5} & \multicolumn{1}{l}{ No.6 } & \multicolumn{1}{l}{No.7 } & \multicolumn{1}{l}{ No.8 }\\
%\multicolumn{1}{l}{Wavelength} & \multicolumn{1}{l}{SRC(Err)} & \multicolumn{1}{l}{ SRC(Err)} & \multicolumn{1}{l}{SRC(Err)} & \multicolumn{1}{l}{ SRC(Err) } &
%\multicolumn{1}{l}{ SRC(Err)} & \multicolumn{1}{l}{ SRC(Err)} & \multicolumn{1}{l}{ SRC(Err)} & \multicolumn{1}{l}{SRC(Err)}\\
           &No.9     &No.10      &No.11     &No.12      &No.13     &No.14      &No.15      &No.16 \\
Wavelength &SRC(Err) &SRC(Err)  &SRC(Err) &SRC(Err)  &SRC(Err) &SRC(Err)  &SRC(Err)  &SRC(Err) \\
\hline \hline
\endfirsthead
\multicolumn{9}{r}%
{{\bfseries \tablename\ \thetable{} -- continued from previous page}} \\
           &No.9     &No.10      &No.11     &No.12      &No.13     &No.14      &No.15      &No.16 \\
Wavelength &SRC(Err) &SRC(Err)  &SRC(Err) &SRC(Err)  &SRC(Err) &SRC(Err)  &SRC(Err)  &SRC(Err) \\
\hline \hline
\endhead
\hline \multicolumn{9}{r}{{Continued on next page}} \\
\endfoot
\endlastfoot

5680 &-0.004(0.088) &0.027(0.015) &0.007(0.033) &0.013(0.005) &0.017(0.049) &0.004(0.000) &-0.013(0.041) &0.024(0.008) \\
5780 &0.251(0.031) &0.198(0.028) &0.041(0.007) &0.027(0.005) &0.065(0.012) &0.045(0.007) &0.086(0.010) &0.085(0.012) \\
5880 &0.508(0.063) &0.340(0.048) &0.159(0.021) &0.136(0.020) &0.251(0.034) &0.183(0.022) &0.254(0.031) &0.275(0.031) \\
5980 &0.665(0.067) &0.437(0.055) &0.330(0.039) &0.329(0.045) &0.482(0.055) &0.370(0.041) &0.416(0.046) &0.447(0.045) \\
6080 &0.749(0.068) &0.517(0.060) &0.501(0.054) &0.524(0.067) &0.630(0.062) &0.551(0.053) &0.556(0.054) &0.552(0.048) \\
6180 &0.814(0.063) &0.590(0.063) &0.633(0.060) &0.656(0.078) &0.731(0.064) &0.691(0.058) &0.666(0.055) &0.643(0.051) \\
6280 &0.873(0.059) &0.659(0.065) &0.726(0.059) &0.732(0.080) &0.808(0.061) &0.790(0.059) &0.751(0.054) &0.722(0.051) \\
6380 &0.942(0.055) &0.726(0.066) &0.807(0.059) &0.811(0.083) &0.873(0.060) &0.872(0.060) &0.826(0.052) &0.801(0.053) \\
6480 &1.010(0.050) &0.794(0.067) &0.876(0.057) &0.877(0.080) &0.935(0.058) &0.945(0.058) &0.895(0.050) &0.876(0.052) \\
6580 &1.074(0.047) &0.865(0.064) &0.945(0.054) &0.939(0.076) &0.995(0.055) &1.017(0.054) &0.965(0.047) &0.946(0.050) \\
6680 &1.138(0.044) &0.938(0.061) &1.016(0.051) &1.013(0.073) &1.056(0.051) &1.087(0.051) &1.037(0.044) &1.019(0.048) \\
6780 &1.193(0.039) &1.012(0.056) &1.087(0.047) &1.084(0.068) &1.112(0.047) &1.153(0.046) &1.110(0.041) &1.088(0.045) \\
6880 &1.244(0.035) &1.085(0.050) &1.155(0.043) &1.156(0.065) &1.164(0.042) &1.217(0.042) &1.177(0.037) &1.153(0.042) \\
6980 &1.281(0.029) &1.154(0.045) &1.213(0.037) &1.214(0.056) &1.207(0.039) &1.272(0.038) &1.239(0.034) &1.210(0.037) \\
7080 &1.315(0.025) &1.217(0.038) &1.267(0.033) &1.267(0.049) &1.242(0.034) &1.319(0.031) &1.294(0.028) &1.261(0.032) \\
7180 &1.337(0.022) &1.271(0.032) &1.312(0.028) &1.317(0.041) &1.270(0.029) &1.357(0.026) &1.338(0.024) &1.307(0.027) \\
7280 &1.347(0.020) &1.314(0.026) &1.344(0.025) &1.354(0.035) &1.293(0.027) &1.385(0.023) &1.371(0.022) &1.342(0.023) \\
7380 &1.350(0.017) &1.345(0.017) &1.369(0.019) &1.383(0.025) &1.315(0.018) &1.405(0.015) &1.392(0.014) &1.368(0.016) \\
7480 &1.345(0.021) &1.363(0.014) &1.382(0.016) &1.402(0.017) &1.333(0.014) &1.416(0.011) &1.399(0.010) &1.383(0.013) \\
7580 &1.331(0.023) &1.369(0.012) &1.386(0.013) &1.409(0.013) &1.337(0.014) &1.412(0.014) &1.394(0.011) &1.386(0.012) \\
7680 &1.307(0.023) &1.365(0.016) &1.381(0.014) &1.407(0.016) &1.333(0.019) &1.397(0.018) &1.379(0.015) &1.377(0.016) \\
7780 &1.271(0.028) &1.353(0.022) &1.366(0.020) &1.393(0.021) &1.320(0.023) &1.368(0.022) &1.353(0.017) &1.360(0.020) \\
7880 &1.227(0.024) &1.333(0.026) &1.344(0.019) &1.367(0.030) &1.298(0.025) &1.330(0.025) &1.322(0.021) &1.333(0.022) \\
7980 &1.180(0.026) &1.309(0.034) &1.319(0.025) &1.338(0.038) &1.267(0.029) &1.285(0.030) &1.283(0.025) &1.300(0.027) \\
8080 &1.131(0.027) &1.282(0.041) &1.290(0.032) &1.296(0.049) &1.228(0.033) &1.236(0.034) &1.243(0.032) &1.262(0.033) \\
8180 &1.078(0.035) &1.254(0.051) &1.255(0.040) &1.244(0.057) &1.185(0.036) &1.183(0.039) &1.202(0.036) &1.220(0.038) \\
8280 &1.024(0.044) &1.227(0.055) &1.215(0.048) &1.183(0.062) &1.140(0.037) &1.130(0.041) &1.160(0.042) &1.176(0.045) \\
8380 &0.982(0.048) &1.200(0.064) &1.179(0.051) &1.137(0.062) &1.096(0.046) &1.080(0.045) &1.120(0.043) &1.141(0.046) \\
8480 &0.937(0.052) &1.164(0.068) &1.140(0.055) &1.090(0.064) &1.054(0.050) &1.033(0.048) &1.079(0.047) &1.102(0.051) \\
8580 &0.888(0.062) &1.121(0.071) &1.093(0.057) &1.046(0.069) &1.013(0.052) &0.989(0.048) &1.039(0.048) &1.055(0.051) \\
8680 &0.828(0.070) &1.070(0.074) &1.040(0.062) &0.992(0.071) &0.969(0.054) &0.945(0.051) &0.994(0.050) &0.998(0.052) \\
8780 &0.756(0.072) &1.005(0.071) &0.971(0.060) &0.936(0.072) &0.911(0.058) &0.892(0.049) &0.931(0.049) &0.933(0.052) \\
8880 &0.672(0.078) &0.920(0.078) &0.883(0.063) &0.854(0.074) &0.828(0.063) &0.815(0.053) &0.851(0.055) &0.853(0.051) \\
8980 &0.562(0.079) &0.787(0.062) &0.754(0.056) &0.716(0.070) &0.700(0.047) &0.693(0.047) &0.720(0.046) &0.734(0.048) \\

\noalign{\smallskip}\hline
\end{longtable}
%\end{landscape}

\clearpage
%\end{CJK}
\end{document}